\documentclass[fleqn,usenatbib]{mnras}

\usepackage[T1]{fontenc}
\usepackage{ae,aecompl}

\usepackage{graphicx}	
\usepackage{amsmath}	
\usepackage{amssymb, multirow}	

\title[38 Eri from the ground and  from space]{
The Delta Scuti star 38 Eri from the ground and from space\thanks{
The ground-based observations were obtained at SAAO, South Africa; UNAM San Pedro Mart\'{\i}r, Mexico; 
Siding Spring, Australia; Zahran, Iran; Sobaeksan, Korea, 
Albuquerque, USA;  Almaty, Kazakhstan and La Palma, Spain.}\thanks{ 
Based on data from the {\it MOST} satellite, a Canadian Space Agency
mission, jointly operated by Dynacon Inc., the University of Toronto
Institute for Aerospace Studies and the University of British
Columbia, with the assistance of the University of Vienna.}
}

\author[M. Papar\'o et al.]
{M. Papar\'o,$^{1}$\thanks{E-mail: {\tt paparo@konkoly.hu}}
Z. Koll\'ath,$^{2}$
R.~R. Shobbrook,$^{3}$\thanks{Dedicated to the memory of R.~R. Shobbrook who passed away during the preparation
of the paper.}
J.~M. Matthews,$^{4}$
V. Antoci,$^{5}$\newauthor
J.~M. Benk\H{o},$^{1}$
N.-K. Park,$^{6,7}$
M.~T. Mirtorabi,$^{8}$
K. Luedeke,$^{9}$ 
A. Kusakin,$^{10}$\newauthor
Zs. Bogn\'ar,$^{1}$  
\'A. S\'odor,$^{1}$
A. Garc\'ia-Hern\'andez,$^{11}$
J.~H. Pe\~na,$^{12}$
R. Kuschnig,$^{13,14}$\newauthor
A.~F.~J. Moffat,$^{15}$
J. Rowe,$^{16}$
S.~M. Rucinski,$^{17}$ 
D. Sasselov,$^{18}$
and W.~W. Weiss$^{13}$
\\
Affiliations are listed at the end of the paper
}

\date{Accepted 2018 April 3. Received 2018 March 26; in original form 2017 November 2}

\pubyear{2018}

\begin{document}
\label{firstpage}
\pagerange{\pageref{firstpage}--\pageref{lastpage}}
\maketitle

\begin{abstract}
We present and discuss the pulsational characteristics of the Delta Scuti star 
38 Eri from photometric data obtained at two widely spaced epochs, partly from 
the ground (1998) and partly from space ({\it MOST}, 2011). We found 18 frequencies
resolving the discrepancy among the previously
published frequencies. Some of the frequencies appeared  with different relative amplitudes 
at two epochs, however, we carried out investigation for amplitude variability for only the {\it MOST} data. 
Amplitude variability was found for one of three frequencies that satisfy the 
necessary frequency criteria for linear-combination or resonant-mode 
coupling. Checking the criteria of beating and  
resonant-mode coupling we excluded them as possible reason for amplitude variability.
The two recently developed methods of rotational-splitting and sequence-search were applied to 
find regular spacings based only on frequencies. Doublets or incomplete multiplets with $l=1, 2$ and 3 
were found in the rotational splitting search. In  
the sequence search method we identified four sequences. 
The averaged spacing, probably  a combination of the large separation and the rotational frequency, is 
$1.724\pm0.092$~d$^{-1}$. Using the spacing and the scaling relation 
$\bar\rho = [0.0394, 0.0554]$~gcm$^{-3}$ was derived. The shift of the sequences proved to 
be the integer multiple of the rotational splitting spacing.  
Using the precise {\it MOST} frequencies and multi-colour photometry in a hybrid way, 
we identified four modes with $l=1$, two modes with $l=2$, two modes with
$l=3$, and two modes as $l=0$ 
radial modes. 

\end{abstract}




\section{Introduction}

Delta Scuti stars are located at the intersection of the classical 
instability strip and the main sequence on the Hertzsprung-Russell diagram (HRD). 
This region represents a transition zone from radiative cores and thick 
convective envelopes (low mass stars) to large convective cores and thin 
convective envelopes (high mass stars).  Evolved $\delta$ Scuti stars, which are  
beyond the main sequence, experience large changes in their interiors in a 
relatively short period of time after the hydrogen in their core is exhausted. 
The pulsation modes exhibit mixed characters, a g-mode character near the core and 
a p-mode character near the surface \citep{Osaki75}. The 
complexity of both the radial and non-radial modes excited in these stars and 
their critical evolutionary stage makes these stars observationally interesting 
and theoretically challenging. The additional effect of generally being 
considered moderate and fast rotators \citep{Breger00a} increases the 
difficulty for unique mode identification. We present here the $\delta$ 
Scuti star 38 Eri, that has proven challenging even observationally, due to the serious one-day alias problem.

\begin{table}
\begin{centering}
\caption[]{Physical parameters of 38 Eri derived by \citet{Balona00}  and \citet{McDonald17}.
Both set of parameters were used to calculate the period ratios and pulsation constant of 
some frequencies.}\label{tab_phys}
\begin{tabular}{@{}lll@{}}
\hline
  & \citet{Balona00} & \citet{McDonald17} \\ 
$\log T_{\mathrm{eff}}$   &  $3.85\pm0.01$ & $3.85\pm0.01$ \\ 
$\log g$ & $3.60\pm0.06$  &  3.514 \\
$M_V$ $\vert$  $M_{\mathrm{bol}}$ & $1.11\pm0.09$~mag & 1.24~mag \\
$\log L/L_{\sun}$ & $1.50\pm0.04$ &  $1.4077\pm0.0006$ \\
$M/M_{\sun}$ & $2.03\pm0.07$ & assumed \\
$R/R_{\sun}$ & $3.7\pm0.1$  & 3.422  \\
$v\sin i$ & 98 kms$^{-1}$ & \\
$P_{\mathrm{rot}}$ & $< 1.9$ day &  $< 1.8$~day \\
Parallax & $25.98\pm1.04$ mas & $26.80\pm0.012$~mas \\
\hline
\end{tabular}
\end{centering}
\end{table}

\begin{table*}
\begin{centering}
\caption[]{List of observing sites and the contribution to the campaign. 
The best coverage was in South Africa and Mexico. Australian contribution in 
two photometric systems are highly valuable.
}\label{tab_sites}
\begin{tabular}{@{}llcccc@{}}
\hline
Site & Telescope & $T_{\mathrm{base}}$ & No. nights & Cov. of total run &  Filter   \\
  & (m) & (d) & (d) & (\%) &        \\
\hline
SAAO, Sutherland, South Africa & 0.5  & 25 & 11 & 26.2 & {\it uvby, I} \\ 
UNAM, San Pedro Mart\'{\i}r, Mexico & 0.84  & 15 & 13 & 27.3 & {\it BVI}  \\
Siding Spring, Australia & 0.6  & 26 & 9 & 6.3 & {\it vby, UBVRI}  \\
IASBS, Zanjan, Iran & 0.35  & 42 & 6 & 2.4 & {\it V}  \\
Sobaeksan Observatory, Korea & 0.61  & 48 & 4 & 2.6 &  {\it y}  \\
Albuquerque, USA & 0.25  & 50 & 6 & 0.8 &  {\it V}  \\
Fesenkov Observatory, Almaty, Kazkhstan & 0.48 &  & 1 & & {\it V}  \\
\hline
\end{tabular}
\end{centering}
\end{table*}

The light variation of 38 Eri ($o^1$ Eri, HR 1298, HD 26574, F2II-III, {\it V}=4.026~mag) was 
discovered in a short-period variability survey for bright stars of spectral 
type A and F in the southern hemisphere by \citet*{Jorgensen71}.
 The 12-hour observing run revealed a frequency at 13.227~d$^{-1}$ and another 
extremely uncertain one at around 10~d$^{-1}$. The next observing run 
\citet{Jorgensen75} corrected the value of the dominant 
mode to 12.27~d$^{-1}$ and a new peak appeared at 7.74~d$^{-1}$. The third 
photometric trial for determining the frequency content better \citep{Poretti89} 
was a one-week long run consisting of altogether 40 hours of observations. 
The highest peak was at 13.38~d$^{-1}$ and a well-defined structure appeared 
around 6~d$^{-1}$ with peaks at 6.94~d$^{-1}$ and 6.03~d$^{-1}$. The author 
argued that the two frequencies were definitely not an alias of each other. 
Some features were better represented in this interpretation but the fit was 
still not satisfactory. To obtain a better fit additional terms (10.45, 11.84. 
6.74 and 8.17~d$^{-1}$ frequencies) were involved, however these frequencies 
were uncertain according to the paper.

Two spectroscopic investigations were devoted to 38 Eri. \citet{Yang86}  
confirmed the presence of non-radial pulsation. At least four different absorption features 
were localised moving through the spectral line from the high precision, but 
short-duration radial velocity measurements. The mean period 11.36~d$^{-1}$ was 
suggested to be a high degree mode, possibly $l=m=14$.
   
\citet{Balona00} presented 247 high-dispersion echelle spectra for 38 Eri. Due to 
the single-site observation the solution was also affected by the $\pm1$~d$^{-1}$ 
alias problem. Frequencies at 11.9 or 12.9; 12.4 or 13.4 and 16.9 or 
17.9~d$^{-1}$ were reported. Significant frequencies were found at 2.68, 
13.26 and 9.47~$d^{-1}$. The probable identifications were pro-grade 
and retrograde modes of $l=7$. \citet{Balona00} derived the 
physical parameters of 38 Eri  from the dedicated high-dispersion (0.06-0.08~{\AA}) 
spectroscopy that we presented here in Table~\ref{tab_phys}. 
His modelling resulted in a post-main sequence evolutionary stage. Over the years new 
parameters were published. A lower surface gravity value ($\log g=3.39$) and a 
similar effective temperature ($\log T_{\mathrm{eff}}=3.84$) were published 
by \citet{Gray06}. However, the actual resolution was closer to 3.5~{\AA} that is perfect for 
the original goal, for the classification purpose of a large sample. The newly 
reduced {\it Hipparcos} measurements revised the parallaxes to 26.80$\pm$0.32~mas \citep{van Leeuwen07}. 
\citet*{McDonald17} based on the {\it Gaia} measurements, published
 the parameters of 38 Eri that we also present in Table~\ref{tab_phys}. 
According to the paper, the surface 
gravity was calculated from assumed mass. We calculated the rotation period 
using their radius and the rotational velocity supposing $i=90^{\circ}$ inclination, similar to \citet{Balona00}. We used both set of values for 
calculating the pulsation constant and period ratios for some frequencies of 
38 Eri.

The present investigation is the next step in resolving the complex pulsational 
behaviour of 38 Eri.

\section{New observations}\label{sec:obs}

\subsection{Ground-based multi-site campaign in 1998}

The only chance in 1998  was a multi-site campaign to resolve the severe $\pm1$~d$^{-1}$ alias problem.
The campaign extended from mid-October to the end of 
December in 1998. Three sites were dominant regarding the time and the amount of the 
data that were devoted to the joint efforts. These three sites gave the 
compact, 
best coverage part of the campaign. Some other sites also joined the campaign 
but with less contribution. However, they extended the time-base of the campaign.
Table~\ref{tab_sites} gives the sites, the telescope size that was used, their time-base, 
and the number of clear skies during the time span. For better comparison the 
coverage was counted for each observing site. The last column contains the filter(s) that were 
used in the measurements.

 The following instruments were applied:
\begin{itemize}
\item
South-African Astronomical Observatory, Sutherland,  South Africa: 
single-channel Modular Photometer with a HAMAMATSU R943-02 Gallium 
Arsenide tube \citep{Kilkenny88} with a
neutral density filter. 
\item UNAM, San Pedro Mart\'{\i}r, Mexico: a single-channel photometer with a 
dry-ice-cooled 1P21 photomultiplier, with a neutral density filter.
\item ANU, Siding Spring Observatory, Australia: the 
`Monitored Filter Box' photometer with an 8-hole filter wheel and computer 
controlled rotations was used. A neutral density filter was used to be able to 
cover both the Str\"omgren and Johnson passbands. 
\item Sobaeksan Observatory, Korea: Photometrics PM512+NU200 
Liquid Nitrogen Cooled CCD Camera coupled with ADPS (Automatic Differential Photometry System).
\item Institute for Advanced Studies in Basic Sciences Observatory (IASBS), 
Zanjan, Iran: SSP-3 photometer was attached to a 14$^{\prime\prime}$ Celestron telescope.
\item Albuquerque, USA: an SSP-3 photometer was attached to a 10-inch Meade 
LX200 telescope
\item Almaty, Kazakhstan: an AZT-14 telescope with an AMF-6 photometer was 
used.
\end{itemize}

\begin{table}
\begin{centering}
\caption[]{The journal of observations. The seven sites are ideally distributed in 
longitude. The total length gives the whole effort independently of colours.
}\label{tab_log}
\begin{tabular}{@{}llc@{\hspace{3pt}}rr@{\hspace{3pt}}c@{}}
\hline
Site & Date & HJD$-$ & Length & N & Observer \\ 
 & 1998 & 2\,450\,000 & (h) &  &  \\
\hline
Siding Spring & 14 Nov & 1132.14 & 2.46 & 16 & RS\\ 
              & 15 Nov & 1133.02 & 1.38 & 10 & RS\\
              & 16 Nov & 1133.95 & 4.83 & 25 & RS\\
              & 20 Nov & 1137.94 & 6.82 & 38 & RS\\
              & 23 Nov & 1140.99 & 4.43 & 17 & RS\\
              & 26 Nov & 1144.05 & 3.75 & 19 & RS\\
              & 29 Nov & 1146.94 & 6.34 & 33 & RS\\
              & 01 Dec & 1148.96 & 6.39 & 33 & RS\\
              & 11 Dec & 1158.07 & 3.17 & 18 & RS\\
Subtotal         &        &         & 39.57 & 209 &\\
\noalign{\smallskip}
San Pedro Mart\'ir & 26 Oct & 1113.79 & 5.33 & 22 &ZK/JP\\
                 & 27 Oct & 1114.79 & 7.11 & 51 & ZK\\
                 & 28 Oct & 1115.74 & 7.20 & 66 & ZK\\
                 & 31 Oct & 1118.72 & 7.11 & 74 & ZK\\
                 & 01 Nov & 1119.72 & 7.38 & 71 & ZK\\
                 & 02 Nov & 1120.77 & 6.46 & 68 & ZK\\
                 & 03 Nov & 1121.71 & 7.33 & 80 & ZK\\
                 & 04 Nov & 1122.72 & 7.11 & 76 & ZK\\
                 & 05 Nov & 1123.72 & 7.02 & 52 & ZK\\
                 & 06 Nov & 1124.73 & 6.99 & 59 & ZK\\
                 & 07 Nov & 1125.86 & 4.25 & 31 & ZK\\
                 & 09 Nov & 1127.72 & 7.39 & 70 & ZK\\
                 & 10 Nov & 1128.72 & 7.38 & 38 & ZK\\
Subtotal            &        &         & 88.06 & 758 &\\
\noalign{\smallskip}
Sutherland & 04 Nov & 1122.31 & 7.31 & 36 & MP\\
           & 07 Nov & 1125.30 & 7.29 & 60 & MP\\
           & 08 Nov & 1126.30 & 7.48 & 62 & MP\\
           & 10 Nov & 1128.31 & 7.36 & 46 & MP\\
           & 11 Nov & 1129.29 & 7.68 & 58 & MP\\
           & 21 Nov & 1139.40 & 4.85 & 43 & MP\\
           & 23 Nov & 1141.27 & 5.46 & 70 & MP\\
           & 24 Nov & 1142.31 & 5.14 & 45 & MP\\
           & 25 Nov & 1143.27 & 4.67 & 43 & MP\\
           & 28 Nov & 1146.32 & 6.06 & 54 & MP\\
           & 29 Nov & 1147.32 & 5.87 & 53 & MP\\
Subtotal      &        &         & 69.17 & 570 &\\
\noalign{\smallskip}
Iran       & 27 Oct & 1114.34 & 3.85 & 36 & TM\\
           & 31 Oct & 1118.31 & 2.78 & 30 & TM\\
           & 03 Nov & 1121.31 & 5.04 & 52 & TM\\
           & 13 Nov & 1130.33 & 3.87 & 42 & TM\\
           & 25 Nov & 1143.26 & 3.89 & 46 & TM\\
           & 08 Dec & 1156.29 & 4.56 & 42 & TM\\
Subtotal   &        &         & 23.99 & 248 &\\
\noalign{\smallskip}
Korea      & 11 Nov & 1129.13 & 4.32 & 37 & NKP\\
           & 28 Nov & 1146.04 & 5.40 & 35 & NKP\\
           & 27 Dec & 1174.94 & 5.88 & 51 & NKP\\
           & 29 Dec & 1176.95 & 3.00 & 27 & NKP\\
Subtotal      &        &         & 18.6 & 150 &\\
\noalign{\smallskip}
USA        & 19 Oct & 1105.79 & 1.99 & 27 & KL\\
           & 05 Nov & 1122.73 & 1.91 & 25 & KL\\
           & 14 Nov & 1131.81 & 1.89 & 27 & KL\\
           & 22 Nov & 1139.81 & 1.90 & 27 & KL\\
           & 24 Nov & 1141.76 & 1.14 & 14 & KL\\
           & 08 Dec & 1155.71 & 0.98 & 14 & KL\\
Subtotal      &        &         & 9.81  & 134 &\\
\noalign{\smallskip}
Kazakhstan  & 08 Dec & 1156.19 & 3.57 & 23 & AK\\
           &        &         &      &     & \\ 
Total      &        &         & 252.77 & 2092 &\\
\hline
\end{tabular}
\end{centering}
\end{table}

Table~\ref{tab_log} lists the journal of the observation with dates on which observations 
were obtained, the time of the beginning of observations in HJD, the duration of 
those observations in hours and decimals, the 
number of measurements of 38 Eri and the name of the observers. Subtotal of the 
observing time and the number of the measurements are given for each site. The 
total values for the whole campaign are given in the last line.

At majority of the sites the
C1= HR 1272 (A1V) and C2= HR 1290 (G8III) were used as comparison stars. 
The observing cycle was usually C1, V, C2, V, C1,\dots.  Regarding the brightness of 
38 Eri at some sites, using larger telescopes, a neutral density 
filter was used. The integration time was rather different on the different 
sites from 2 sec (de-focused CCD) to 30-40 sec (neutral filter), due to the 
different 
instruments, filters and weather conditions. In order to assign similar weight, the 
integrations were averaged to get a sampling as dense as at other sites. The 
sampling times were 6-7 minutes at most sites. Siding Spring observatory has a 
slightly longer sampling time (7-9 minutes) due to the longer observing cycles using most
passbands of two photometric systems (Str\"omgren and Johnson). This contribution 
turned out to be highly important in the mode identification.

If the observing run was long 
enough, the nightly determined extinction coefficients were used. In other cases 
mean values were applied. If some trend remained it was subtracted as a 
straight line. The V$-$C1 (38 Eri $-$ HR 1272) relative light curve was used for 
the frequency analyses. The C1$-$C2 curve (HR 1272 $-$ HR 1290) was used for checking 
the non-variability of the C1= HR 1272 comparison star. In a single night only 
C2= HR 1290 was observed. 
The V$-$C1 curve was determined using the literature value of the C1$-$C2.

Different photometer/filter combinations lead to different zero points, even if 
the same comparison stars were used. This fact needs to be considered when 
different data sets are combined.
The differential light curves showed that the zero points of the instrumental 
system on the different sites were slightly different. All light curves were 
shifted to the zero point by subtracting the nightly mean value.

Combining the data of each site in a certain colour resulted in the campaign data 
for 1998. Table~\ref{tab_sites} shows that both Str\"omgren {\it vby} and 
Johnson-Cousins {\it BVI} measurements
 were obtained at least at two sites. The observations of R. Shobbrook at 
Siding Springs were connected partly to the Str\"omgren observations in South 
Africa and the Johnson {\it BVI} observation in Mexico.

\subsection {\textit{MOST} data from 2011}

The {\it MOST} (Microvariability and Oscillation of STars) Canadian Space Telescope 
\citep{Walker03} was launched on June 30, 2003 into a low-Earth (altitude 
820 km), Sun-synchronous circular polar orbit with an orbital period of 101~minutes. 
The orbital period corresponded to an orbital frequency of 14.2~d$^{-1}$.
The light was collected by a 15-cm aperture  Rumak-Maksutov telescope to two
frame-transferred CCD cameras, through a single, wide passband filter 
(350-700~nm). One CCD camera was planned only for tracking the telescope, while other was used 
for scientific purposes. After dramatically reducing the tracking jitter in 2005, the {\it MOST}
telescope simultaneously supplied three types of photometric data for 8 weeks in the 
Continuous Viewing Zone (CVZ) of declination range of $+34 > \delta > -18$.

In the Fabry imaging mode the bright stars illuminate the CCD through a Fabry 
jet lens. The Direct Imaging mode resembles conventional CCD photometry where
photometry is obtained from slightly defocused images of stars. In the Guide Star 
Photometry mode the {\it MOST} Star tracker (the originally non-scientific CCD camera)
 was used for photometry. The satellite operations were upgraded after the 
launch, so the Attitude Control System (ACS) helped in providing more highly accurate photometry 
(see, e.g. \citealt{Walker05,Aerts06}).

\begin{figure}
\includegraphics[scale=0.34]{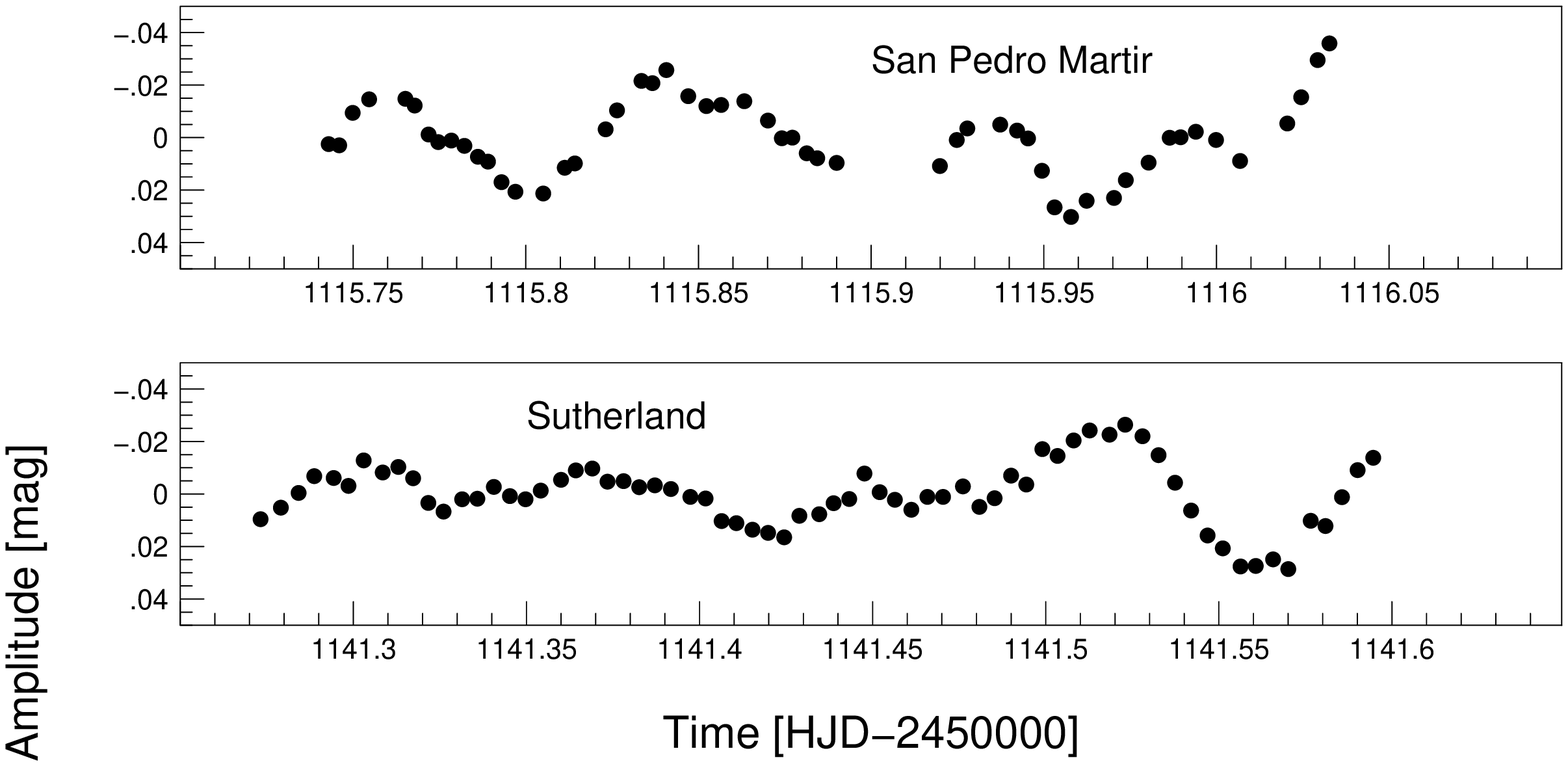}
\includegraphics[scale=0.34]{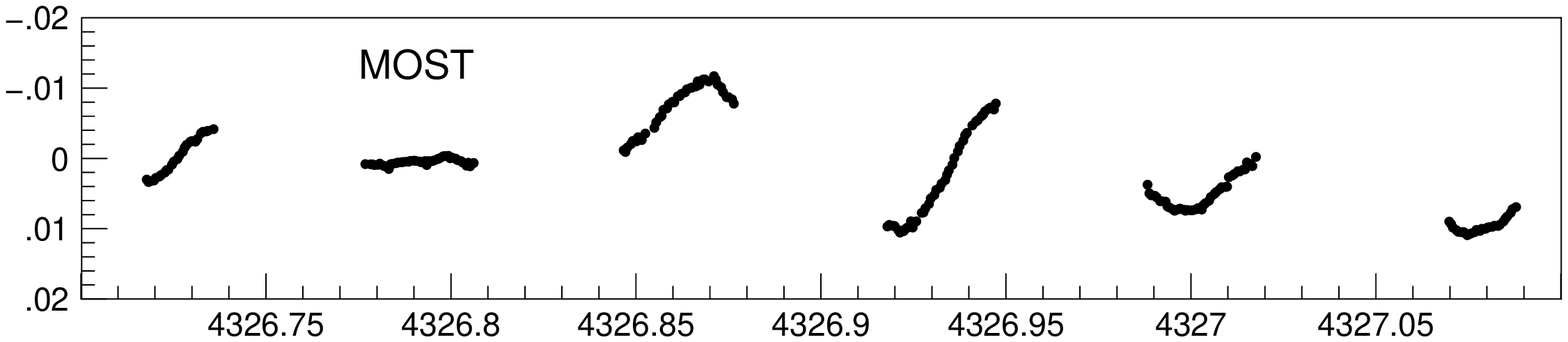}
\includegraphics[scale=0.34]{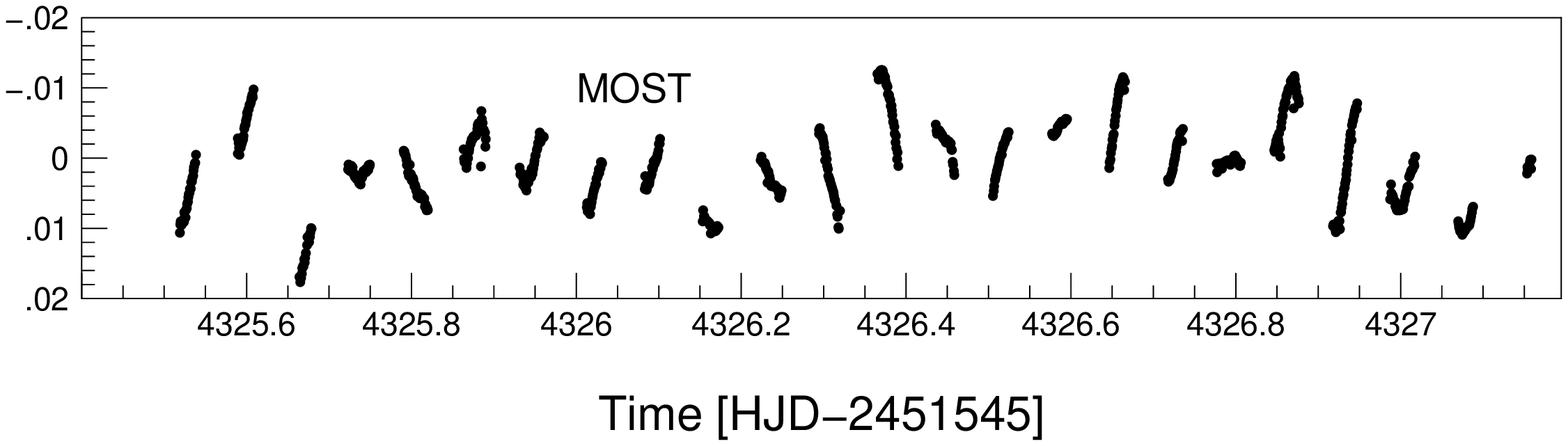}
\caption{
Light curves at San Pedro Mart\'{\i}r, South-Africa from ground and by {\it MOST} from space.
The upper three panels show the quality of the ground-based data and the space photometry on the same scale. 
The regular gaps in the {\it MOST} data are shown in the bottom panel.
}
\label{light}
\end{figure}

{\it MOST} was designed to achieve the mission's primary goal of detecting rapid 
photometric oscillations (period of several minutes) in bright pulsating stars 
with a precision of approximately 1 parts per million ($\mu$mags) in the Fourier domain. 
{\it MOST} was intended for non-differential photometery, since the relatively high frequencies of
the oscillations can be clearly distinguished in the Fourier spectrum of the 
data from possible drifts and noise \citep{Matthews04a}.
However, the {\it MOST} space-based photometer could also monitor bright $\delta$ Scuti 
stars, with slower oscillation than the original goal, with an amplitude 
detection limit of a few $\mu$mag. \citet{Matthews04a} predicted (based on a 
rotational simulation  on {\it MOST} data) that the rotational splitting of  
$\delta$ Scuti stars is so complex that even the improved resolution due to the long and continuous time base 
($\approx 0.2\mu$Hz equals to 0.01728 d$^{-1}$) and 
sensitivity (ppms) possible from space observations do not necessarily yield unambiguous
result. Since 38 Eri with its $\delta_{2000}=-6^{\circ} 50^{\prime} 16^{{\prime\prime}}$ position fits the 
{\it MOST} Viewing Zone, we hoped that the {\it MOST} observations could help to solve the
ambiguities of the frequencies previously published.
 
The target, 38 Eri was observed by {\it MOST} from Nov 05, 2011 to Dec 04, 2011 in the Fabry imaging mode. 
It shared each {\it MOST} orbit with two other Prime Science targets, so there were gaps spaced 
by the {\it MOST} orbital period of about 101 minutes. The individual exposure was 
3.0-second long, but stacked for a sampling cadence of once per 30 seconds. 
The modulation of scattered Earth shine at the orbital period of the {\it MOST}
satellite was filtered by
subtracting a linear fit (`sky' versus star) from the data. This method shows 
the best result and modulates the stellar signal least but gets rid of the 
orbital period quite nicely. The reduced data presented by the {\it MOST} team 
contains 
HJD = HJD$_{\mathrm{star}}-$HJD$_\mathrm{{MOST}}$  
where HJD$_\mathrm{{MOST}} = 51\,545.0$, the standard 
zero-point epoch of {\it MOST}. In the second column the brightness as
mag$=-2.5\log$(ADU pix$^{-1}$ sec$^{-1}$) is given normalised to the mean. The 
light curve has the time base of 29.5 days 
(Rayleigh resolution is 0.034~d$^{-1}$) and 11\,187 measurements.
A sample of the most compact ground-based light curves and  two subsets 
of the {\it MOST} light curve are given in Figure~\ref{light}. The upper three panels 
show the quality of the ground-based data at San Pedro Mart\'ir (Mexico) and 
Sutherland (South-Africa), and the space photometry by {\it MOST} on the same scale. 
The bottom panel shows the regular gaps in the 
{\it MOST} data over the orbital period. Obviously, the space data are superior to 
the ground-based observations despite the gaps
\footnote{All photometric data obtained from the ground and from space are given in the electronic tables
attached to this paper.}.

\subsection {Spectroscopy from 2013}

We obtained 69 high-resolution spectra with the HERMES spectrograph at the 
1.2-m Mercator Telescope on La Palma \citep{Raskin11}, Canary Islands, Spain 
between 14 and 23 January, 2013. The 
fiber-fed echelle spectrograph was operated in high-resolution (HRF) mode, 
providing $R=85\,000$ resolving power. The spectra were reduced with the dedicated 
reduction pipeline of the HERMES instrument, including bias and stray-light 
subtraction, echelle-order extraction, flat-field correction, cosmics 
filtering, and wavelength calibration using ThAr wavelength reference spectra 
which were obtained several times during the night \citep{Raskin11}. 
Afterwards, we normalised the spectra employing an in-house developed automatic 
tool by \'A.S. based on expected continuum wavelength-sections relying on a 
synthetic F0 stellar spectrum. Finally, we cross-correlated the spectra with a 
theoretical F0-type spectrum of Solar metallicity without line broadening.

\begin{table}
\begin{centering}
\caption[]{Colour observations. The longest data sets were obtained 
in Johnson {\it V} and Str\"omgren {\it y}. The fourth column gives the 
coverage for the whole data set and in some colours for the most compact part, too.
}\label{tab_color}
\begin{tabular}{@{}lrrcc@{}}
\hline
Filter & Length & N & Coverage & Observer \\ 
 & (h) &  & (\%) &        \\
\hline
{\it u} & 74.70 & 535 & 12.0 & MP  \\
{\it v} & 107.20 & 722 & 12.1/16.5 & MP,RS  \\
{\it b} & 113.53 & 803 & 17.5 & MP,RS \\ 
{\it y} & 122.82 & 820 & 9.3 & MP,RS,NKP  \\
{\it U} & 32.90 & 131 & 5.5/9.1 &  RS  \\
{\it B} & 107.18 & 800 & 9.9/12.8 & ZK,RS \\
{\it V} & 143.60 & 1030 & 11.7/17.1 & ZK,RS,TM,KL,AK \\
{\it R} & 29.83 & 131 & 5.0/9.1 & RS \\
{\it I} & 76.50 & 765 & 14.0/19.1 & MP,RS,ZK \\
\hline
\end{tabular}
\end{centering}
\end{table}

\section{Frequency analysis}\label{sec:data}

\subsection{Spectroscopic analysis}

The 69 high-resolution spectra obtained in a one-week long interval were not
adequate enough in order to find the frequencies. Nevertheless, we 
derived the $v\sin i$ and the dominant 
frequency from the data at hand.

\subsubsection{Projected rotational velocity}

\begin{figure}
\begin{center}
\includegraphics[scale=0.95]{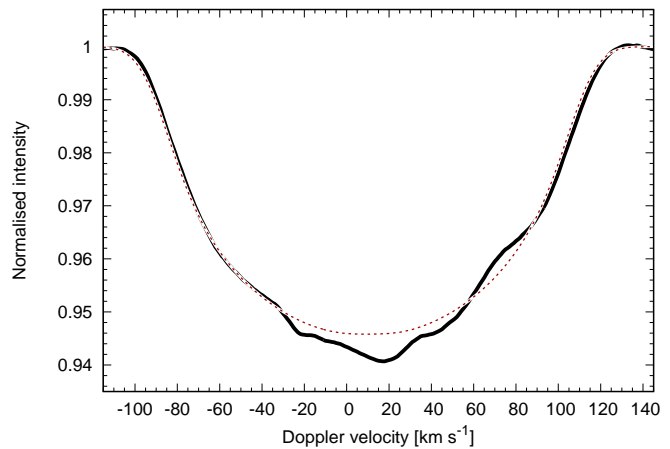}
\end{center}
\caption{The mean CCF of 38 Eri (thick black line) and the fitted rotationally 
broadened line profile (thin dashed curve).}
\label{fig:vsini}
\end{figure}

We used the {\sc Famias} software \citep{Zima08a, Zima08b} to fit a rotationally 
broadened absorption-line profile to our mean cross-correlation function (CCF) in order to derive 
$v\sin i$ of 38~Eri. The best-fit model has a 
$v\sin i = 101$~km\,s$^{-1}$ projected rotational velocity and 
$\sigma = 10$~km\,s$^{-1}$ intrinsic line-width. The mean CCF along 
with the fit is plotted in Fig.~\ref{fig:vsini}.

Note the strong asymmetry and irregularities of the average CCF. This is most
probably, due to the low number (69) of spectroscopic observations which are not enough to average out 
the complex multi-periodic pulsational variations from 
the mean profile. Furthermore, the pulsational line-profile variations, as a 
kind of macro-turbulence, contribute to the line-broadening. Thus, the real 
$v\sin i$ value might be below the obtained value of 101~km\,s$^{-1}$. 
Unfortunately, the low number of spectra does not allow us to model the 
pulsational line-profile variations; therefore, the precise amount of this 
contribution is unknown. Practically our solution fits the previously 
determined $v\sin i$, namely 98~km\,s$^{-1}$ by \citet{Baglin73} and \citet{Balona00}.

\subsubsection{Frequency analysis from spectroscopy}

\begin{figure}
\begin{center}
\includegraphics[angle=0,scale=0.73]{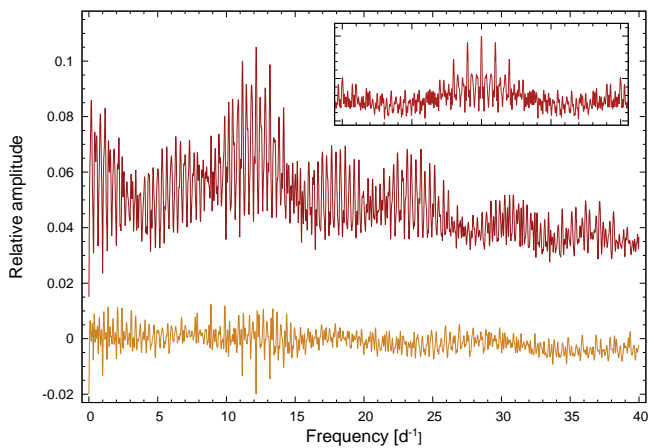}
\end{center}
\caption{Fourier pixel-by-pixel spectrum of the CCFs of 38 Eri. 
Top: original spectrum. Bottom: spectrum pre-whitened by the significant 
variation of 12.16~d$^{-1}$ (shifted by $-0.02$ in relative amplitude for better visibility).
Insert shows the window function.}
\label{fig:pbpsp}
\end{figure}

We investigated the line-profile variations in the 
CCFs by Fourier-analysis using the pixel-by-pixel method, and also with 
{\sc Famias} \citep{Zima08a,Zima08b}. Fig.~\ref{fig:pbpsp} shows the Fourier 
spectrum of the CCF variations averaged over the dispersion range $-80$ to 
$+110$~km\,s$^{-1}$. Note that the exact choice of the dispersion range does not 
affect the overall result. The highest peak is at 12.16~d$^{-1}$. Our observations 
were affected by daily aliases, as the spectral window function in the 
insert of Fig.~\ref{fig:pbpsp} demonstrates. Therefore, the real variation 
frequency might be $\pm 1$~d$^{-1}$ off from this value. The bottom function in 
Fig.~\ref{fig:pbpsp} shows the residual spectrum after the 12.16~d$^{-1}$ variation 
had been pre-whitened from the data. Due to the low number of our spectroscopic 
observations, no further significant line-profile variation frequencies could be 
identified.

\subsection{Frequency analyses of photometric data}

The multi-frequency analysis of 38 Eri was performed with the {\sc MuFrAn} program 
\citep{Kollath90}. {\sc MuFrAn} (MUlti FRequency ANalysis) is a collection of 
frequency determination, sine fitting for observational data and graphics 
routines for the visualisation of the results. At each step all previous 
frequencies were computed.
{\sc Period04} was only used for getting the variability of the given frequency. 
{\sc MuFrAn} does not have this option. The two programs, tested during the work 
on the first {\it CoRoT} RR Lyrae star \citep{Chadid10}, are completely equivalent 
concerning the results on the frequencies.

\begin{figure*}
\includegraphics[angle=0,scale=0.6]{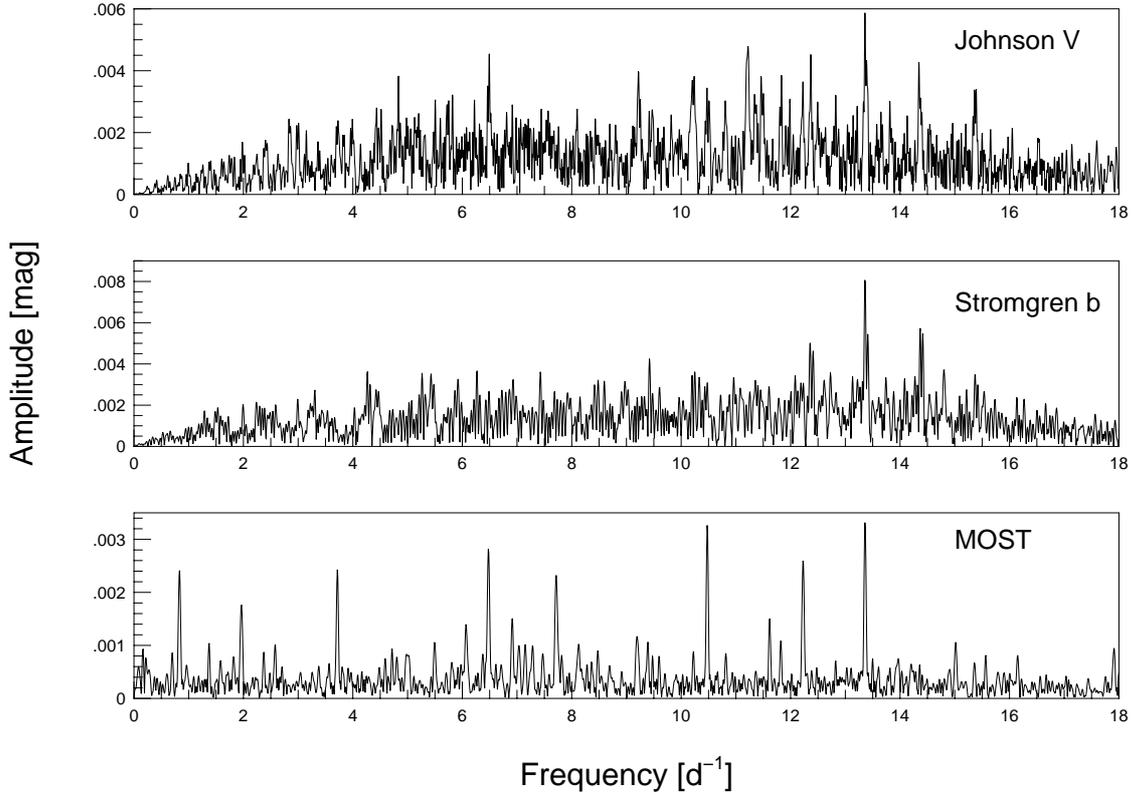}
\caption{
Fourier spectrum in Johnson {\it V}, Str\"omgren {\it b} colours and 
{\it MOST} data. The first two panels show how different the frequency 
content is in the different passbands, although the spectral windows are similar 
(see time base and coverage in Tables~\ref{tab_sites} and \ref{tab_color}. 
The {\it MOST} spectrum is much cleaner; the peaks are more pronounced due to the 
higher precision and better data distribution. 
The peaks below 4~d$^{-1}$ disappear in the frequency extraction procedure.
}
\label{spec}
\end{figure*}

The periodicity search was carried out for both the ground-based and {\it MOST} data.
The ground-based campaign provided a differential light curve of V$-$C1, which 
was analysed for the frequencies. After checking the frequency content of C1$-$C2 
differential light curve we concluded that every peak in the 
Fourier spectrum of the V$-$C1 differential light curve is attributed to the 
variability of 38 Eri.

The {\it MOST} data present a non-differential light variation of 38 Eri. 
The peaks of the Fourier spectrum are attributed to the light variation of 38 Eri, although 
aliases due to the orbital period appear. Peaks in the low-frequency region 
may appear as an alias pattern caused by the orbital period (linear combination 
of the excited modes and the orbital period). 

Figure~\ref{spec} contains the Fourier spectra of the ground-based data (in Johnson {\it V} 
and Str\"omgren {\it b} colours) and the {\it MOST} data. Despite the 
17 percent coverage of the 
ground-based colour data (that is half of the {\it MOST} coverage), 
the original Fourier spectrum 
shows that the effect of the 1~d$^{-1}$ alias in the ground based data was much more 
severe than the alias of the orbital period in the {\it MOST} Fourier spectrum. We 
expected that the variability of 38 Eri is much better reproduced by the 
frequencies obtained by {\it MOST}. We can also see at first glance from the two panels 
that the frequency content is rather different in the different passbands. The 
obviously present peak at around 6.5~d$^{-1}$ in Johnson {\it V} band does not seem 
to be present/dominant in the Str\"omgren {\it b} band. There were so different 
solutions in the 
previous analyses of 38 Eri at different epochs, and in some sense the 
ground-based campaign and 
the {\it MOST} data also resulted in different solutions that we present both analyses 
in a bit more detail.

\subsubsection{Frequencies in the ground-based campaign}

\begin{figure*}
\includegraphics[angle=0,scale=0.6]{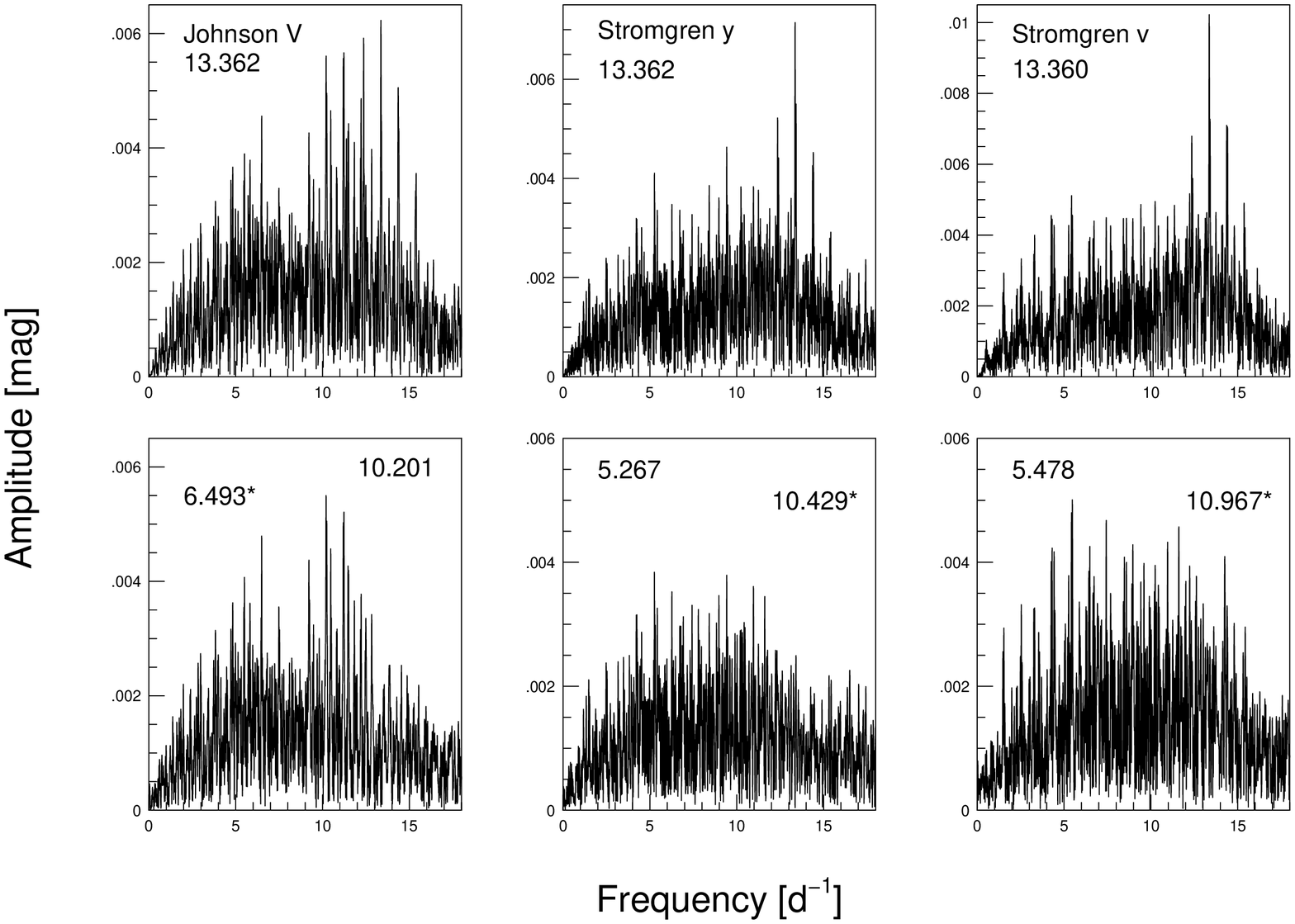}
\caption{
Fourier spectra in Johnson {\it V}, Str\"omgren {\it y} and {\it v} colours.
The panels in the upper row show that the same frequency is dominant in each passband. 
In the bottom row the first steps of the pre-whitening process are given.
One of the labels gives the peak of the highest amplitude in the spectrum.
The numbers with * mean the third frequencies of the highest amplitudes 
in the pre-whitening process.
}
\label{spect}
\end{figure*}

The observation in different colours provided the possibility to obtain the 
frequencies from data of different time-length, thus producing different spectral 
windows. The resolution of frequencies is lower from the shorter time span. 
However, the more compact data distribution produces a less complicated spectral window, 
which means less alias problem. Table~\ref{tab_color} gives details on the colour 
observation: the filter, the observing length in hours and decimals, the number of 
measurements, the coverage for all and for the compact part, as well as the observers.

\begin{table*}
\begin{centering}
\caption[]{Frequencies in Johnson-Cousins colours. The frequency solutions were 
independently obtained. Beside the frequencies and amplitudes the links to 
the {\it MOST} solution are given. The Rayleigh resolution at the ground-based 
data is 0.028~$d^{-1}$. 
}\label{tab_fr_Johnson}
\begin{tabular}{@{}rrcrrcrrc@{}}
\hline
\multicolumn{3}{c}{Johnson {\it B}} & \multicolumn{3}{c}{Johnson {\it V}} & \multicolumn{3}{c}{Cousins {\it I}} \\ 
ID & Frequency & Amp & ID & Frequency & Amp & ID & Frequency & Amp   \\ 
 & (d$^{-1}$) & (mmag) &  & (d$^{-1}$) & (mmag) &  & (d$^{-1}$) & (mmag)     \\
\hline
$f_1$ & 13.364 & 9.58 & $f_1$ & 13.362 & 6.12 & $f_1$ & 13.359 & 4.61  \\
$f_{18}$ & 10.214 & 7.41 & $f_{18}$ & 10.201 & 5.43 & $f_5$ & 11.622 & 2.08  \\  
$f_2$ & 10.481 & 4.99 & $f_3$ & 6.493 & 5.46 & $f_3+1$ & 7.436 & 2.52  \\
$f_3$ & 6.499 & 5.02 & $f_{17}$ & 11.832 & 4.10 & $f_{12}$ & 7.786 & 1.70 \\
$f_{17}+1$ & 12.836 & 2.78  & $f_2$ & 10.473 & 3.47 & $f_{17}+1$ & 12.819 & 1.57 \\
$f_8-1$ & 8.107 & 4.68  & $f_7-1$ & 4.988 & 4.48 & $f_8$ & 9.018 & 1.79  \\
$f_{15}$  & 5.782  & 4.22 & $f_{15}+1$ & 6.767 & 4.23 & $f_9$ & 7.033 & 0.55     \\
$f_{12}$ & 7.883 & 3.49 & $f_{12}$ & 7.634 & 2.99 & & &    \\
$f_{15}-1$  & 4.743 & 3.80 & $f_{14}-1$ & 9.817 & 3.08 & & &    \\
$f_2+1$ & 11.420 & 2.67 &  & 5.201 & 2.68 & & &  \\
\hline
\end{tabular}
\end{centering}
\end{table*}

\begin{table*}
\begin{centering}
\caption[]{Frequencies in Str\"omgren colours. Links to the {\it MOST} solution are given. 
No link means that those frequencies were not found in the {\it MOST} data. 
The most remarkable case is 5.267~d$^{-1}$ in Str\"omgren {\it y} colour. 
The Rayleigh resolution at the ground-based data is 0.028~d$^{-1}$.
}\label{tab_fr_Stromgren}
\begin{tabular}{@{}rrrrrrrrrrrr@{}}
\hline
\multicolumn{3}{c}{Str\"omgren {\it u}} & \multicolumn{3}{c}{Str\"omgren {\it b}} & \multicolumn{3}{c}{Str\"omgren {\it v}}  & \multicolumn{3}{c}{Str\"omgren {\it y}} \\
ID & Freq & Amp & ID & Freq & Amp & ID & Freq & Amp & ID & Freq & Amp   \\ 
 & (d$^{-1}$) & (mmag) &  & (d$^{-1}$) & (mmag) &  & (d$^{-1}$) & (mmag) & & (d$^{-1}$) & (mmag)     \\
\hline
$f_1$ & 13.363 & 10.48 & $f_1$ & 13.360 & 8.56 & $f_1$ & 13.360 & 11.45 & $f_1$ & 13.362 & 6.26 \\
$f_3-1$ & 5.433 & 3.90 & $f_2$ & 10.477 & 3.22 & $f_3-1$ & 5.478 & 5.85 &  & 5.267 & 3.14 \\  
$f_5+1$ & 12.624 & 4.54 & $f_{18}$ & 10.204 & 3.16 & $f_{14}$ & 10.967 & 4.62 & $f_2$ & 10.429 & 5.47  \\
$f_6$ & 6.920 & 4.34 & $f_3-1$ & 5.429 & 3.13 & $f_{12}-1$ & 6.728 & 3.60 & $f_3+1$ & 7.424 & 3.54  \\
 & 5.267 & 3.37  & $f_5$ & 11.609 & 3.55 &  & 6.254 & 3.50 & $f_8$ & 8.974 & 2.46 \\
$f_2$ & 10.475  & 3.51  & $f_6$ & 6.924 & 3.72 & $f_5+1$ &  12.573 & 3.44 & $f_5$ & 11.376 & 3.91  \\
$f_{14}$  & 10.962  & 2.44 &  & 14.258  & 3.10 &  & 14.248 & 3.37 & $f_4$ & 12.236 & 2.98 \\
$f_{16}+1$ & 14.924 & 2.80 &  & 13.233 & 2.73 & $f_2$ & 10.475 & 3.96 & & 10.206  & 2.81   \\
$f_3+1$  & 7.412  & 2.84 &  & 6.264 & 2.88 & $f_7$ & 6.089 & 3.53 & & &   \\
 & 13.278  & 2.23 &  &  &  & $f_{13}$ & 9.601 & 3.10 & & &  \\
$f_8+1$  & 10.191  & 1.98 &  &  &  &  &  &  & & &   \\
\hline
\end{tabular}
\end{centering}
\end{table*}

We performed the frequency analysis independently for each data set, for 
different observers and summing up for different colours. The independent analysis meant 
that the consecutive frequency of the largest amplitude was picked up. We did 
not have an `a priori' frequency solution in our mind from the previously published 
frequencies or from the analyses of the previous data sets. 
Although the length of the data in different passbands and the Rayleigh resolution 
were the same (0.028~d$^{-1}$) the analyses resulted in different sets of frequencies. However,  
the dominant frequency was definite at 
around 13.36~d$^{-1}$ in each passband, something similar to what \citet{Poretti89} arrived at from 
photometry and \citet{Balona00} from spectroscopy. There was a 
$-1$~d$^{-1}$ alias ambiguity in \citet{Balona00}'s result.

The consecutive 2-3 steps showed a colour-dependent diversity. 
We do not present all steps of the frequency search process in each colour here. In 
Figure~\ref{spect} we show only some representative Fourier spectra, where we can follow 
the different frequencies appearing in the second and third steps. The third 
frequency is labelled by * after the frequency's value in each panel.

In Johnson {\it V} the second peak of the highest amplitude is 10.201~d$^{-1}$,
that 
has never been reported before. We do not force the explanation that
it could be the $-2$~d$^{-1}$ alias of 12.27~d$^{-1}$ reported by \citet{Jorgensen75}.
The third peak at 6.493~d$^{-1}$ has not been reported either. Only a 
well-defined structure around 6~d$^{-1}$ with a peak at 6.94 and 6.03~d$^{-1}$ 
was mentioned by \citet{Poretti89}.

In Str\"omgren {\it y} colour also a new, previously not reported peak appeared at 
5.267~d$^{-1}$ as a second peak. The third peak with nearly the same amplitude was 10.429~d$^{-1}$, that is 
close to the 10.45~d$^{-1}$ uncertain solution given by \citet{Poretti89}. 
Surprisingly the third peak is almost at  twice the value of the second peak.
A joint analysis of Johnson {\it V} and Str\"omgren {\it y} light curves were also 
carried out. Beside the dominant peak at 13.36~d$^{-1}$, the second and third frequencies are 
10.201~d$^{-1}$ and 10.481~d$^{-1}$ in the frequency search process.

In Str\"omgren {\it v} colour the second peak was at 5.478~d$^{-1}$ which could be the
$-1$~d$^{-1}$ alias of the second peak found in Johnson {\it V} colour. However,
the third peak at 10.967 is also almost twice the value of the second peak.
If we compare the 2.68~d$^{-1}$ poorly determined frequency given by \citet{Balona00} 
to the 5.478 and 10.967~d$^{-1}$ frequencies, we are reminded of a well-known 
process of chaotic behaviour, period bifurcation. Of course, the
ground-based data are not as high quality for having such a strong conclusion, 
but it is worthwhile to mention this surprising numerology. 

For providing a complete view on how different the frequencies in the different colour bands are,
we present the complete frequency list of the ground-based color observations.
Table~\ref{tab_fr_Johnson} and Table~\ref{tab_fr_Stromgren} contain the 
frequencies obtained in the Johnson-Cousins {\it BVI} and 
Str\"omgren {\it uvby} colours, respectively. The frequencies are listed according 
to decreasing amplitude. Beside the frequencies and amplitudes,
the tables show a link to the finally accepted {\it MOST} frequencies given later in 
Table~\ref{tab_fr_MOST}, and discussed in Sec~\ref{sec:comp}. 

\subsubsection{Frequencies in MOST}

\begin{figure*}
\includegraphics[angle=0,scale=0.75]{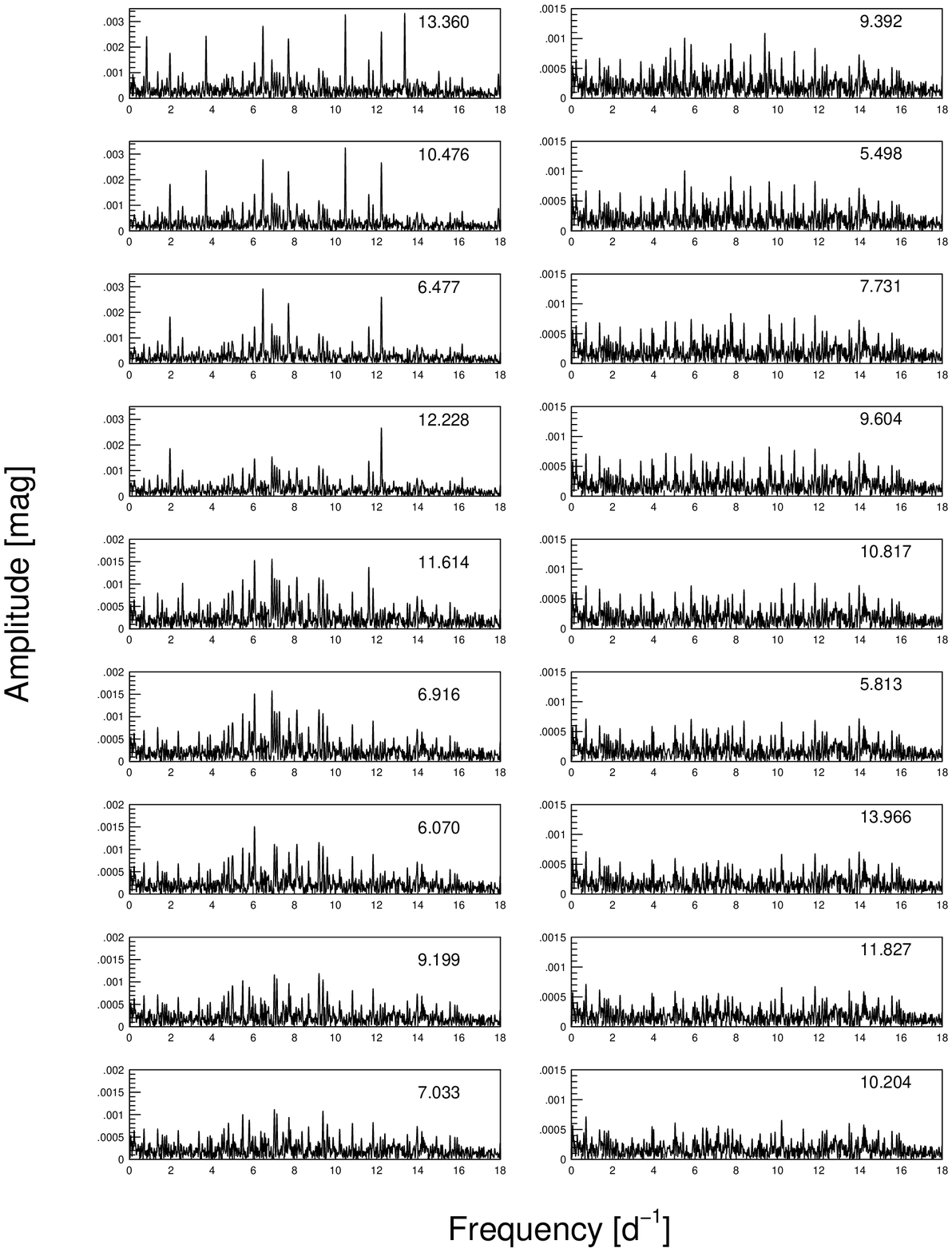}
\caption{
Steps of frequecy search in the Fourier analyses of {\it MOST} data. 
The amplitude scale was changed from the fifth and the tenth panel for getting 
a better view of the low amplitude peaks. The label marks the peak of the highest amplitude.
}
\label{prew}
\end{figure*}

The diversity of the ground-based results shows that 
it was worth updating the ground-based campaign data with the {\it MOST} observations.
The original spectrum in Figure~\ref{spec} shows how much more definite the frequency 
content of the {\it MOST} data for 38 Eri was. There is no $+1$ or $-1$~d$^{-1}$ alias and 
the final frequency list shows why the single-site data suffered from  
severe alias problem. Figure~\ref{prew} shows the steps of the frequency search process of 
the {\it MOST} data. The labels provide the exact frequency value of the dominant peak 
in the panel. 

\begin{table}
\begin{centering}
\caption[]{Frequency solution of {\it MOST} data. Frequencies are given both 
in d$^{-1}$ and $\mu$Hz. In the phase calculation HJD 2\,451\,546.0 epoch was used. 
The errors of the amplitudes and phases are calculated by {\sc Period04} \citep{Lenz05}. 
}\label{tab_fr_MOST}
\begin{tabular}{@{}rrrlc@{}}
\hline
ID & Frequency & Frequency & Amp & Phase   \\ 
 & (d$^{-1}$) & $(\mu$Hz) & (mmag) & (rad)         \\
\hline
\noalign{\smallskip}
$f_1$ & 13.360 & 154.629 & 3.38$\pm0.07$ & 3.387$\pm0.018$ \\
$f_2$ & 10.476 & 121.250 & 3.37$\pm0.07$ & 4.008$\pm0.021$ \\  
$f_3$ & 6.477 & 74.965 & 2.98$\pm0.08$   & 4.917$\pm0.027$ \\
$f_4$ & 12.228 & 141.527 & 2.73$\pm0.06$ & 5.921$\pm0.020$ \\
$f_5$ & 11.614 & 134.421 & 1.37$\pm0.07$ & 0.615$\pm0.058$ \\
$f_6$ & 6.916 & 80.046 & 1.57$\pm0.08$  & 6.068$\pm0.052$  \\
$f_7$ & 6.070 & 70.255 & 1.56$\pm0.05$ & 1.686$\pm0.038$   \\
$f_8$ & 9.199 & 106.470 & 1.22$\pm0.05$ & 5.809$\pm0.048$  \\
$f_9$ & 7.033 & 81.400 & 1.13$\pm0.08$ & 6.120$\pm0.073$   \\
$f_{10}$ & 9.392 & 108.704 & 1.03$\pm0.07$ & 1.485$\pm0.057$ \\
$f_{11}$ & 5.498 & 63.634 & 0.95$\pm0.10$ & 4.115$\pm0.087$  \\
$f_{12}$ & 7.731 & 89.479 & 1.30:$\pm0.07$ & 4.354$\pm0.032$ \\
$f_{13}$ & 9.604 & 111.157 & 0.88$\pm0.12$ & 3.580$\pm0.075$ \\
$f_{14}$ & 10.817 & 125.197 & 0.68$\pm0.08$ & 0.152$\pm0.090$ \\
$f_{15}$ & 5.813 & 67.280 & 0.73$\pm0.07$ & 2.583$\pm0.127$  \\
$f_{16}$ & 13.966 & 161.644 & 0.70$\pm0.05$ & 2.097$\pm0.079$ \\
$f_{17}$ & 11.827 & 136.887 & 0.71$\pm0.09$ & 5.131$\pm0.115$ \\
$f_{18}$ & 10.204 & 118.102 & 0.70$\pm0.05$ & 4.620$\pm0.083$ \\
\hline
\end{tabular}
\end{centering}
\end{table}

Not only was the dominant frequency at 13.36~d$^{-1}$ confirmed but the second and 
third of the highest amplitudes at 10.476 and 6.477~d$^{-1}$ were also 
definite. The fourth frequency at 12.228~d$^{-1}$ can be detected without any doubt. 
The dominant mode and this latter frequency caused confusion in the \citet{Jorgensen71} and \citet{Jorgensen75}
papers and in the paper of \citet{Balona00}. The consecutive steps
revealed frequencies at 6.916 and 6.070~d$^{-1}$, which are similar to the 
frequencies found by \citet{Poretti89}. 
The possible explanation that the 5.4 and 6.4~d$^{-1}$
 frequencies are 1~d$^{-1}$ aliases was excluded here. 
Beside the frequency at $f_3=6.477$~d$^{-1}$, the other frequency at $f_{11}=5.498$~d$^{-1}$ 
(2nd right panel) also appeared with lower amplitude.
This is another reason for a severe $\pm 1$~d$^{-1}$ alias. However, the 
frequency at 5.2~d$^{-1}$ was not found here in the {\it MOST} data. 
The other frequency that shows relatively much lower amplitude is 
$f_{18}=10.201$~d$^{-1}$ that was the frequency of the second highest amplitude 
in Johnson {\it B} and {\it V} colours.

\begin{figure*}
\includegraphics[width=\columnwidth]{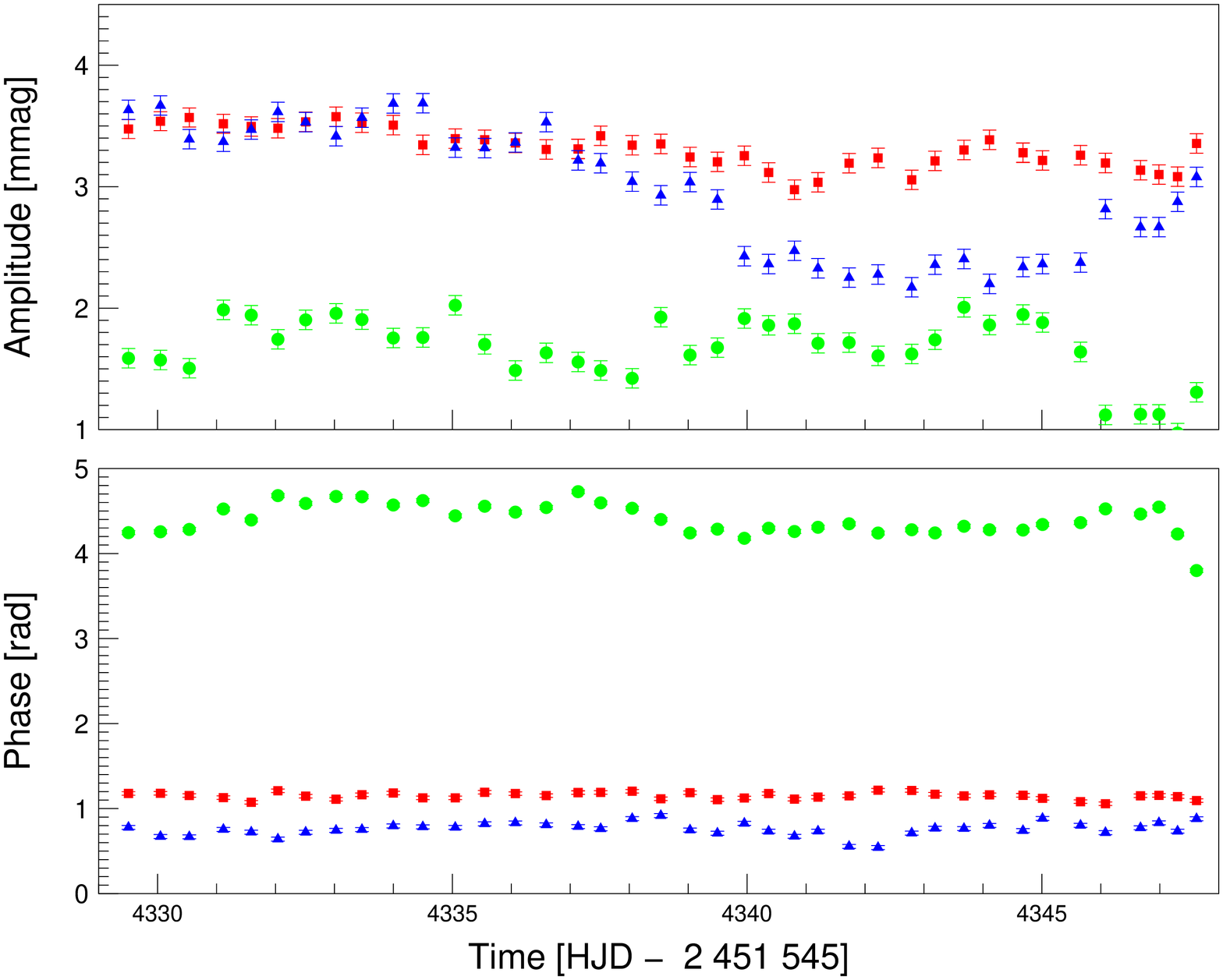}
\includegraphics[width=\columnwidth]{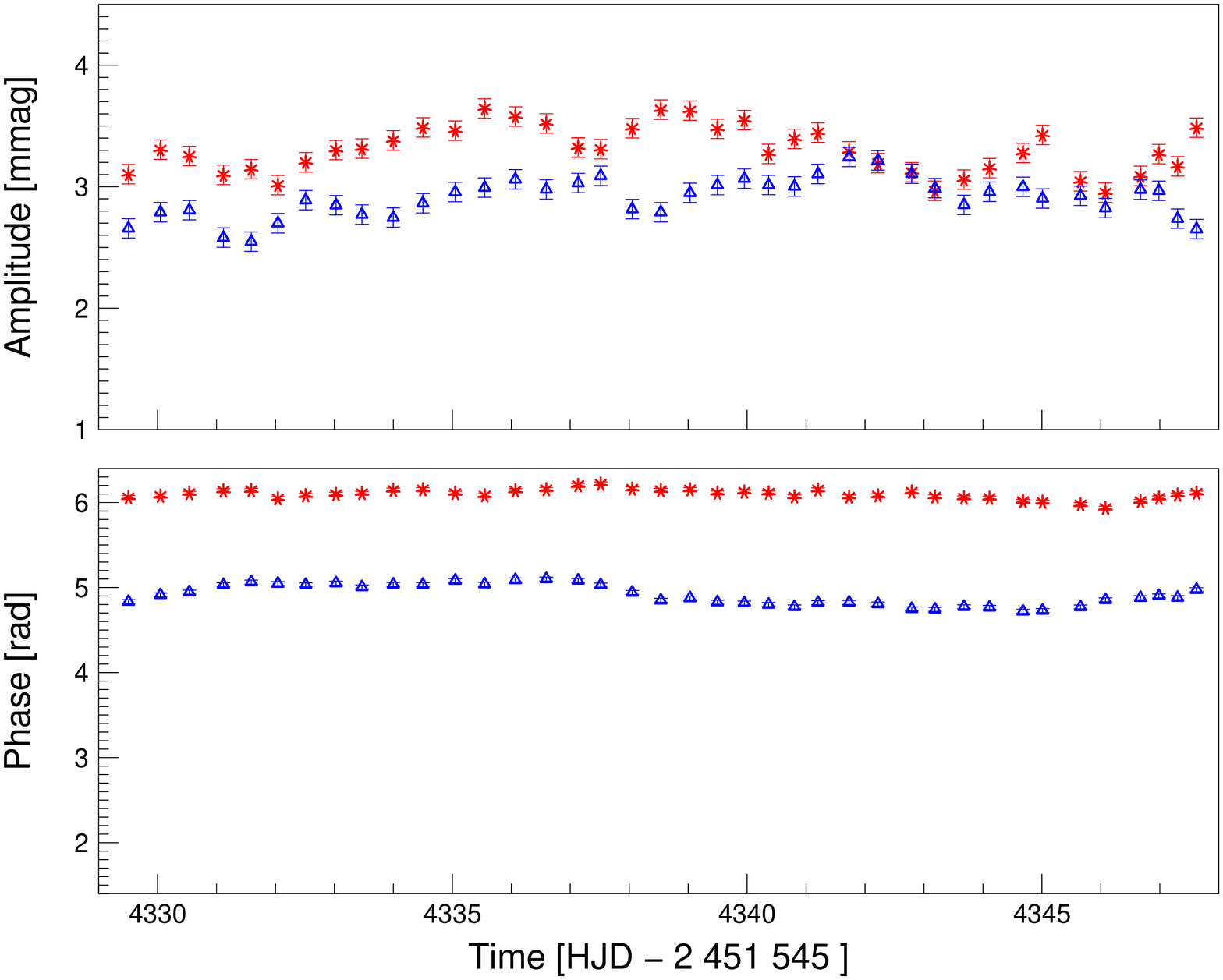}
 \caption{The amplitude variation of five high
amplitude frequencies: (left) $f_1$ (red squares), $f_3$
(blue triangles),
$f_6$ (green dots), (right) $f_2$ (red asterisks),
$f_4$ (blue open triangles). The corresponding panels are at the same
scales.
}
\label{fig:amp_var}
\end{figure*}

As we see in the {\it MOST} data set the $\pm 1$~d$^{-1}$ aliases are resolved. However, in the 
{\it MOST} spectrum an alias structure appears at the orbital frequency at 
14.2~d$^{-1}$. An interesting effect is that the sum of the frequencies at 
6.477 and 7.731~d$^{-1}$ is equal to 14.2~d$^{-1}$, the orbital 
frequency. Pre-whitening with the higher amplitude frequency at 6.477~d$^{-1}$, 
the rather high amplitude peak at 7.7~d$^{-1}$ disappears, but after many steps 
we were able to localise a peak at $f_{12}=7.731$~d$^{-1}$ (3rd right panel). However, 
pre-whitening with both
frequencies the amplitude of the latter one increased showing that they could not
 be properly determined (marked by colon in Table~\ref{tab_fr_MOST}). The question arises, whether it is only a 
remnant of the frequency subtracting or it is a real frequency excited in the 
star. The latter version is supported by the fact that a frequency at 
7.74~d$^{-1}$ appeared in the analyses of \citet{Jorgensen75}  and in our 
ground-based data sets,  
that are not connected to the {\it MOST} orbital frequency.
Of course, the most definite answer could be obtained from a continuous data, however, 
unfortunately 38 Eri was not continuously observed by {\it MOST}. 
In some cases an algorithm for filling the gaps and getting continuous data could 
be an acceptable solution.
 \citet*{Pascual-Granado15} developed {\sc Miarma}, a new gap-filling algorithm, 
which is suggested to use for space data with gaps smaller than the data 
segments (like gaps in {\it CoRoT} light curves due to the South Atlantic Anomaly). 
However, predictions lose coherence rapidly when gaps are much larger than the 
data segments. We could not use {\sc Miarma} for the shared {\it MOST} data, 
since the gaps are longer than the data segments.

We also found a numerical relation between the dominant mode (13.360~d$^{-1}$) 
and the sum of 6.477 and 6.916~d$^{-1}$ (equal to 13.393~d$^{-1}$). Regarding
the 0.034~d$^{-1}$ resolution of the {\it MOST} data, the agreement is remarkable. 
This nearly same value to the dominant mode with the lower resolution of the 
ground-base data increased the confusion around the dominant mode of 38 Eri. 
The {\it MOST} data set with its better resolution can resolve these three 
frequencies without any doubt.

The finally accepted list of frequencies in 38 Eri are given in Table~\ref{tab_fr_MOST}.
The table contains the  ID of the frequencies, the frequencies both in 
d$^{-1}$ and $\mu$Hz, the amplitude and the phases to the {\it MOST} epoch (HJD$-$2\,451\,545.0).
These IDs are linked to the ground-based frequencies in Table~\ref{tab_fr_Johnson} 
and Table~\ref{tab_fr_Stromgren}.

\subsubsection{Comparison of the ground-based and {\it MOST} frequencies}\label{sec:comp}

The links in Table~\ref{tab_fr_Johnson}
and Table~\ref{tab_fr_Stromgren} remind us of the serious $\pm 1$~d$^{-1}$ alias problem of the
ground-based observation, even though it was a multi-site international campaign.
In many cases the frequencies obtained on the colour data could be connected to 
the $+1$ or $-1$~d$^{-1}$ alias of the {\it MOST} frequencies. 
However, there are frequencies in Table~\ref{tab_fr_Stromgren} (Str\"omgren colours) 
where no link to the {\it MOST} frequencies is given. 
The most interesting examples are 5.26~d$^{-1}$ in {\it u} 
and {\it y} and probably its $+1$~d$^{-1}$ alias at 6.26~d$^{-1}$ in {\it b} and {\it v} colours. 
Both frequencies in each colour, except in {\it y}, appear together with 
5.4 d$^{-1}$,  which could be $f_{11}$ or the possible $-1$~d$^{-1}$ alias of $f_3$ in the list 
of {\it MOST} frequencies. In the Str\"omgren {\it y} colour the 
$+1$~d$^{-1}$ alias of $f_3$ frequency appears. The differences in decimals,
disregarding the $\pm 1$ alias, are 0.166, 0.165, 0.244 and 0.269~d$^{-1}$, which
are larger than the resolution in the colour data sets.
Frequencies at 5.2 and 5.4 d$^{-1}$ seems to represent a pair in the ground-based data.
A similarly close frequency pair appears in Johnson {\it B, V} and Str\"omgren {\it u} at
10.214/10.481, 10.201/10.473 and 10.191/10.475~d$^{-1}$, respectively. The
differences are 0.267, 0.272 and 0.284~d$^{-1}$. Although these frequencies also appear
in the {\it MOST} frequencies as $f_2$ and $f_{18}$, the later one has much lower amplitude 
than the frequencies around 10.2 d $^{-1}$ in the ground-based data. We may speculate whether 
these pairs represent rotational doublets with
changing amplitude at different epochs.

Slightly smaller 
frequency differences appear in decimals between the dominant frequency and
a peak of lower amplitude, namely 13.36/13.278, 13.360/13.233 and 13.360/14.248~d$^{-1}$ 
in Str\"omgren {\it u, b} and {\it v} colours.
These are 0.087, 0.127, 0.113 and 0.083~d$^{-1}$, which are closer to the resolution. 
These frequencies are critical 
in the {\it MOST} data due to the 14.2~d$^{-1}$ orbital frequency.
The frequency solution in the different colour bands would suggest two 
possible explanations: at different epochs different frequencies are excited
with different amplitudes or with the different passbands we are able to catch
the pulsation in structurally changing layers.

The comparison of the 
results of the individual ground-based observing runs and the {\it MOST} run reveal
that the pulsation is rather 
complex and we could see only a certain aspect of the complex pulsation at a certain moment. 
With 18 frequencies we still could not get a perfect fit (residual is 0.0025~mag). The light curve 
contains abrupt changes from extremely large amplitude cycles to cycles with 
very tiny amplitude (see Fig.~\ref{light}). Normal beating of far-away frequencies do not show such behaviour, 
only the interaction of frequencies that are rather close to each other.

\subsection{Amplitude variability of \textit{MOST} frequencies}

The available theoretical models, in a lack of non-linear treatment of 
non-radial pulsation, do not provide any prediction for the amplitude and 
phase variability of the excited modes. Nevertheless, the investigations of 
amplitude and phase variability nowadays are very common, thanks to the 
extended ground-based follow up \citep{Breger09a,Breger10} and the space data of long 
time base ({\it MOST}, \citealt{Walker03,Matthews04b}, 
{\it CoRoT}, \citealt{Baglin06} and {\it Kepler}, \citealt{Burucki10,Koch10}).
A possible amplitude and phase variability could help to explain the frequency spectrum 
that is rather complicated in some cases. 
 Amplitude variability has been published in the wide region of the HRD in the 
last years; for B stars interpreted as rotational modulation \citep{Balona15}, 
for $\delta$ Scuti stars due to resonant mode coupling 
\citep{Barcelo Forteza15}, for roAp star on a shorter time-scale that was 
expected \citep{Medupe15} and even for a white dwarf as an observational 
consequence of a significantly crystallized stellar interior \citep{Hermes15}. 
For more examples we refer for the review of \citet{Guzik16}.
In a recent paper using the {\it Kepler} 
targets \citet{Bowman16} presented an excellent summary on the possible 
cause of the amplitude and phase variation in $\delta$ Scuti stars. 

Although our data do not stretch over as long a time-base as 
long as the {\it Kepler} data, however, the {\it MOST} observations of one month 
time-base gave a chance to investigate the possible amplitude and phase 
variability of at least the frequencies of highest amplitude in our list.

The original decision to present only the first four frequencies was evident, since they 
represent the same amplitude level in the range of 3.38-2.73~mmag in the {\it MOST}
data. However, the numerical relation among the frequencies 
($f_1$, $f_3$ and $f_6$, discussed later) resulted the necessity of presenting the 
amplitude and phase of five frequencies. The 29.5~day long time-base was divided into 
eight-day long segments shifted by
0.5~days to follow the variability as much as possible, although the points are not 
independent. {\sc Period04} \citep{Lenz05} was used to get the amplitude and phase values and the uncertainties 
for the segments. Figure~\ref{fig:amp_var} shows the amplitude (top panel) and phase 
(bottom panel) variability of five frequencies. The error bars represent similar uncertainties that are 
given in Table~\ref{tab_fr_MOST}.
The time base shown in Figure~\ref{fig:amp_var} is shorter than the duration
of the {\it MOST} observation. We used only 
the solution for the complete eight-day long segments. In addition, due to the 
larger gaps in the data much larger error bars resulted at the end, which we omitted from the panel.  
We present three frequencies ($f_1=13.360$, $f_3=6.477$, 
and $f_6=6.916$~d$^{-1}$) in the left side panels and two frequencies ($f_2=10.476$ and 
$f_4=12.228$~d$^{-1}$) in the right panels. 

The constant phases 
show that the not too long time base, therefore not too high resolution, allows
the determination of the possible variability. We regarded the amplitudes of $f_2$ 
and $f_4$ in the right side panel as constant amplitudes based on two criteria. 
First, a numerical test on the data generated 
with four sinus waves on the time distribution of the {\it MOST} data, resulted
in a 
5 percent scatter around the constant amplitude and phase, due to the gaped data 
distribution. The amplitude variation of $f_2$ and $f_4$ were not larger than 5 percent 
around the mean level. Secondly, following the criteria of constancy given by 
\citet{Bowman16}, namely, a frequency exhibits significant amplitude 
variability if at least half of its amplitude bins deviate more than $\pm 5\sigma$ 
from its mean value we concluded that $f_2=10.476$ and $f_4=12.228$~d$^{-1}$ do not show amplitude 
variability. 

The three frequencies on the left side are separately presented, because 
$f_1 \approx f_3+f_6$, which means a necessary criteria for the linear combination or 
resonant mode coupling. However, the phase curves are constant and an 
amplitude variability higher than the $5\sigma$ criteria is valid only for 
$f_3=6.477$~d$^{-1}$. We have three possibly connected frequencies, but only a 
single amplitude variation.

According to \citet{Bowman16} some amplitude variation can be explained by 
beating of two close frequencies \citep{Breger09b} or the mode coupling mechanism \citep{Breger00b}.
\citet{Breger02} showed that pairs of close-frequency modes were found 
near the expected frequencies of radial modes. In the 
mode identification section, the $f_3=6.477$~d$^{-1}$ frequency is a possible 
candidate for being a radial mode according to the period ratio of two {\it MOST} 
frequencies, discussed later.

Both \citet{Breger02} and \citet{Bowman16} used the 0.01~d$^{-1}$ 
value as a criterion for the frequency separation of the two closely spaced 
frequencies. Due to the resolution of our {\it MOST} data (0.034~d$^{-1}$), we could not 
resolve two frequencies as closely spaced as the criterion. We could not exclude 
the cause of the beating for the amplitude variation of $f_3=6.477$~d$^{-1}$. 
If we take the amplitude variation 
as a periodic one with
a $\approx 20$~day period, the necessary frequency difference between $f_3$ and the 
unresolved frequency is $\approx 0.05$~d$^{-1}$. 

In principle, we should be able to resolve such a close frequency to 
$f_3=6.477$~d$^{-1}$. Instead of a close pair we found a frequency at 6.2 (or 5.2) d$^{-1}$ appearing in the 
ground-based data, but not in the {\it MOST} data. 
Such a far-away frequency does not result in amplitude variation. 
In addition, the beating of a pair of pulsation 
mode frequencies, that are close and resolved, appears as periodic amplitude modulation with a 
characteristic sharp change in phase at the epoch of minimum amplitude for 
each frequency in the pair \citep{Breger06}. However, we have 
a constant phase, so we conclude that the amplitude change of $f_3=6.477$~d$^{-1}$ is 
not caused by beating.

The non-linearity and mode coupling are worthwhile to check, because three 
frequencies satisfy the resonance criteria for both the combination frequency and the mode coupling.
\begin{equation}
     f_1 \approx f_3 + f_6
\end{equation}
where $f_1$ is the child mode and $f_3$ and $f_6$ are the parent modes.
The difference between $f_1$ and the sum is 0.03~d$^{-1}$ that is near to our resolution, 
but the criterion is satisfied. According to the model of \citet{Breger14}, the
amplitude of the child mode is a product of the two modes as
\begin{equation}
  A_1 = \mu_\mathrm{c} A_3 A_6 
\end{equation}
in our case, and the phase relation is
\begin{equation}
\phi_1=\phi_3\pm\phi_6.
\end{equation}
Using $A_1=3.38$, $A_3=2.98$, and $A_6=1.57$~mmags from Table~\ref{tab_fr_MOST}, 
we obtained $\mu_\mathrm{c}=0.722$.

The value of $\mu_\mathrm{c}$ characterizes the type of 
connection between the frequencies. If, $\mu_\mathrm{c}\le 0.01$ this implies weak coupling 
and favours a non-linear distortion model producing a combination frequency. 
A value $\mu_\mathrm{c}> 0.1$ implies a strong coupling and favours resonant mode coupling. 
However, in the mode coupling hypothesis, it is required that all three members 
of a family are variable in amplitude, so that the child mode can be 
identified. In our case, only one parent mode exhibits amplitude change. In
addition, $\phi_1=3.387$, $\phi_3=4.917$, and $\phi_6=6.068$ radians do not fulfil the 
phase relation of the resonant coupling. In this family hypothesis, only one 
frequency has variability and only in amplitude. These arguments exclude  
that resonant mode coupling is going on in 38 Eri. What remains for possible explanation?

Although the sum of $f_3$=6.477 and $f_{12}$=7.731~d$^{-1}$ equals to the {\it MOST}
orbital frequency, $f_{\mathrm{orb}}=14.2$~d$^{-1}$, this connection can only slightly modify 
the amplitudes if they are determined at the same time, but it can not cause an 
amplitude variability.
\citet{Bowman16} conjectured for the pure amplitude 
modulation that it could be caused by variable driving and/or damping within the star.

\subsection{Regularity between the frequencies}

We know two facts for rotating $\delta$ Scuti stars in general. They 
pulsate in many non-radial modes. Due to the rotation the non-radial 
modes split in the simplest cases into equidistant triplets or multiplets. 
The first-order effect \citep{Ledoux51}, the second-order effect 
\citep{Vorontson81, Vorontson83, Dziembowski92}, and the 
third-order effect \citep*{Soufi98} were investigated theoretically in the frame
of the perturbative theory. Later, investigation of the intermediate and fast 
rotating stars, using a different approach, were carried out 
\citep*{Lignieres06, Roxburgh06, Lignieres08, Lignieres09, Lignieres10, Reese08, Reese09}.  
In recent years the frequency pattern of the fast rotating stars combined
with theoretical investigation were used to find stellar density 
\citep{Suarez14, Garcia Hernandez15}, and accurate surface gravity
\citep{Garcia Hernandez17}. The frequency regularities help in the multi-colour
mode identification \citep{Reese17} and can be characterized by few parameters,
for example $\delta\nu$ \citep{Moya17},
which might play a role in the seismic indices for $\delta$ Scuti stars 
\citep{Michel17}. 

The search for regular spacing in $\delta$ Scuti stars has a long history,
too, starting from the histogram method \citep{Breger99} to the Fourier 
transform method used at first for ground based data \citep{Handler97}, later 
for the space data producing echelle diagrams 
\citep{Garcia Hernandez09, Garcia Hernandez13, Garcia Hernandez15}.
 
Two dedicated approaches appeared in the last two years. 
Both of them looked not only for a spacing value, but also sequences of regular spacing 
in different region of spacing. One of them originally
concentrates on the distribution of the overtones with a certain $l$ value and 
the shift of the sequences with different $l$ values \citep{Paparo13, Paparo16a, Paparo16b}. 
The existence of the sequences with similar spacing 
(in the 1.5-3.6~d$^{-1}$ region) allowing a tolerance level has been confirmed \citep{Paparo16b}. 
However, in some cases the pure sequence of the modes with the same 
$l$ value are influenced by the rotation split frequencies 
(see the test case of FG Vir in \citealt{Paparo16a}).

The other approach searches for regularities caused by rotational splitting 
\citep{Chen16,Chen17b,Chen17a}. The final goal of all investigation on the 
regular spacing(s) is the same: to solve the problem of mode identification 
in $\delta$ Scuti stars, to reach the level of asteroseismology for stars pulsating 
in the non-asymptotic regime. We present cases here at the first time where 
the new observables derived from the two methods are compared, looking for possible higher 
level regularities.

\subsubsection{Rotation splitting}

\begin{figure}
\includegraphics[angle=0,scale=0.34]{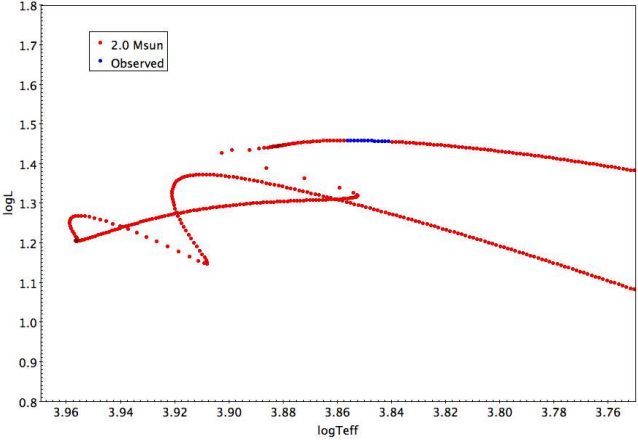}
\caption{
Evolutionary status of 38 Eri.
HR diagram of an evolutionary track with $M = 2$~M$_{\sun}$. 
Red dots show the complete track from the pre-main sequence to
the hydrogen-shell burning phase. Blue dots correspond to the models in
the range of observed $T_{\mathrm{eff}}$ and $\log g$.
}
\label{fig:hrd}
\end{figure}

\begin{figure}
\includegraphics[angle=0,scale=0.34]{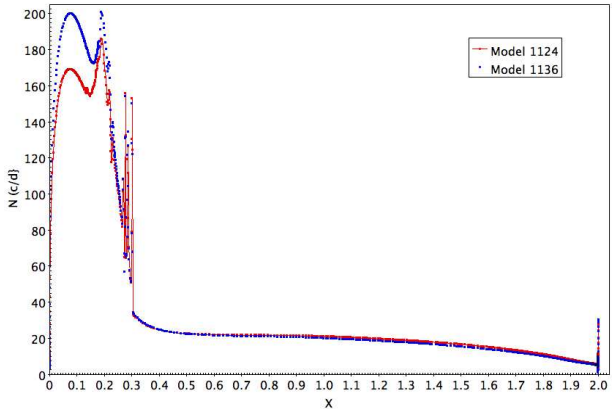}
\caption{
Lamb and Brunt-V\"ais\"al\"a frequencies of 38 Eri.
Brunt-V\"ais\"al\"a frequency ($N$) versus mass for the
limiting models in the range of the observed $T_{\mathrm{eff}}$ and 
$\log g$. Red dots correspond to the less evolved model and blue dots
correspond to the most evolved one. Both models show $N$ values in the core
well above the observed frequencies.
}
\label{fig:NX}
\end{figure}

According to the modelling \citep{Balona00}, 38 Eri is in a shell hydrogen 
burning stage of evolution. This star fits the group of evolved stars 
(EE Cam, HD 50844 and CoRoT 102749568), where the approximate formulae for the 
high-order g-modes 
(derived by \citealt{Brickhill75} and \citealt{Winget91} for white dwarfs) are applied 
for the $\delta$ Scuti stars' mixed modes, that 
is, for modes showing p-mode character in the envelope and g-mode character 
in the inner region. White dwarfs are pulsating in g-modes in the envelope, 
while the g-modes in $\delta$ Scuti stars have low amplitude in the 
envelope.  Although \citet{Dziembowski92} mentioned that g-mode 
asymptotics are relevant also for $\delta$ Scuti stars, we checked
the validity of eq. 4 for 38 Eri. We computed one non-rotating evolutionary track using
the {\sc MESA} code \citep{Paxton11,Paxton13,Paxton15}. We used solar metallicity and mixing
length parameter $\alpha_{\mathrm{MLT}} = 0.5$ following the presciption of
\citet{Garcia Hernandez09} for $\delta$ Scuti stars. The track was computed
with $M = 2$~M$_{\sun}$ \citep{Balona00} and was evolved up to reaching the observed
surface gravity. All the models fitting the observed parameters (Fig.~\ref{fig:hrd}) showed 
a cut-off frequency above 3000~$\mu$Hz ($=259.2$~d$^{-1}$) and a Brunt-V\"ais\"al\"a frequency 
above 160~d$^{-1}$ in the convective core (Fig.~\ref{fig:NX}). The latter value demonstrates 
that 38 Eri fulfills the conditions to apply
eq. 4, i. e., that the rotational frequency is lower than the observed frequencies and
that these are lower than the Lamb and Brunt-V\"ais\"al\"a frequencies.

We applied the 
rotational splitting method for 38 Eri. 
The approximate formula of the 
rotational splitting $\delta\nu_{l,n}$ and the rotational period ($P_\mathrm{{rot}}$) 
for the high order g-modes is 
\begin{equation}\label{eq:nu}
m\delta\nu_{l,m}
=\frac{m}{P_\mathrm{{rot}}}\left[ 1-\frac{1}{l(l+1)}\right]
\end{equation}
where $l, n$, and $m$ are the quantum numbers of the pulsation modes. 
We did not apply the rotational splitting method in the original way. We 
did not search for the 1-20~$\mu$Hz (0.0864-1.728~d$^{-1}$) region 
as \citet{Chen16,Chen17b,Chen17a} did, 
but we did only a visual search among the frequencies (only 18), but all of the small spacings were 
calculated. We noticed that nearby frequencies have
a spacing around 0.2~d$^{-1}$ or its multiple value. These values
were not exactly the same in each case (we may not expect it), so we could
not pick up a certain value for the rotatinal splitting. However, we recognezed
that the most frequent smallest unit of the spacing is near the expected rotational 
splitting, if we suppose that the inclination angle is close to 90$^{\circ}$ and we use 
$P_{\mathrm{rot}}$ obtained from the spectroscopic $v\sin i$. According to our assumption, 
if the inclination is really close to 90$^{\circ}$, then we find spacing with 
these and it proves that our assumption is not completely wrong.
The rotational period  according to \citet{Balona00} and \citet{McDonald17} are
$P_\mathrm{{rot}} < 1.9$  and < 1.8 day for 38 Eri (see in Table~\ref{tab_phys}). 
The expected rotational splittings
according to the formula are $\delta\nu_1>0.263(0.278)$~d$^{-1}$ for $l=1$, 
$\delta\nu_2>0.438(0.463)$~d$^{-1}$ for $l=2$, 
and $\delta\nu_3>0.482(0.509)$ for $l=3$. 
The expected ratios are 
$\delta\nu_1/\delta\nu_2=0.6$, $\delta\nu_3/\delta\nu_2=1.1$ which confirms 
the relation derived by \citet{Winget91}, namely, $\delta\nu_{n,l=1}:
\delta\nu_{n,l=2}: \delta\nu_{n,l=3} = 0.6:1:1.1$. We searched for triplets
and multiplets with these spacings among the frequencies given in Table~\ref{tab_fr_MOST}.

In accordance with the small number of frequencies, it is not surprising
that we did not find clear triplets as in the published papers, 
but we found only some doublets and incomplete multiplets. 
The possible rotational splittings are given in Table~\ref{tab_rot}.

\begin{table}
\begin{centering}
\caption[]{Possible rotational splittings. The spacing values were searched according 
to Eq.~\ref{eq:nu}.  
}\label{tab_rot}
\begin{tabular}{@{}llllll@{}}
\hline
Multiplet & ID & Freq & $\delta\nu$ & $l$ & $m$  \\ 
 &  &  (d$^{-1}$) & (d$^{-1}$) &  &      \\
\hline
\multirow{2}{*}{1} & $f_{15}$ & 5.813 &  \multirow{ 2}{*}{0.257} & 1  & $-1$/0 \\
 & $f_7$ & 6.070 &  & 1  & 0/$+1$ \\
\noalign{\bigskip}
\multirow{2}{*}{2} & $f_2$ & 10.476 &  \multirow{ 2}{*}{0.272} & 1  & $-1$/0  \\
 & $f_{18}$ & 10.204 &  & 1 & 0/$+1$  \\
\noalign{\bigskip}
\multirow{3}{*}{3} & $f_3$ & 6.476 & \multirow{ 2}{*}{0.439} & 2 & $-1$/0    \\
 & $f_6$ & 6.916 & \multirow{ 2}{*}{$2*0.442$} & 2 & 0/$+1$  \\
 & $f_{12}$ & 7.731 &  & 2 & $+2$ \\
\noalign{\bigskip}
\multirow{2}{*}{4} & $f_2$ & 10.476 & \multirow{ 2}{*}{$2*0.436$} & 2 &  $-2$ \\
 & $f_{13}$ & 9.604 &  & 2 & 0   \\
\noalign{\bigskip}
\multirow{2}{*}{5} & $f_{11}$ & 5.498 &  \multirow{2}{*}{$2*0.490$} & 3 &  $-2$/0 \\
 & $f_3$ & 6.477 &  & 3 & 0/$+2$   \\
\noalign{\bigskip}
\multirow{2}{*}{6} & $f_9$ & 7.033 & \multirow{2}{*}{$2*0.482$} & 3 &  $-2$/0 \\
 & $f_7$ & 6.070 &  & 3 & 0/$+2$   \\
\hline
\end{tabular}
\end{centering}
\end{table}

\begin{table}
\begin{centering}
\caption[]{Sequence search result. Four sequences were identified with near similar spacing. 
The averaged spacings are given in the last column. The frequencies are given in d$^{-1}$.
}\label{tab_ssa}
\begin{tabular}{@{}lcrrccc@{}}
\hline
No & Freq & Freq & Freq & Freq & Freq & Av. spacing \\
\hline
I & 5.498 & 6.916 & 10.476 & 12.228 & 13.966 & 1.672 \\
II & 5.813 & 7.731 & 9.604 & 11.614 & 13.360 & 1.887 \\
III & 6.477 & 10.204 & 11.827 &  & & 1.744 \\
IV & 6.070 & 9.199 & 10.817 & & & 1.591 \\ 
\hline
\end{tabular}
\end{centering}
\end{table}

\begin{table*}
\begin{centering}
\caption[]{Multi-colour amplitude ratios and phase differences given 
for mode identification. The averaged values of the phase differences are in the last column. 
The deviations from the average are given in brackets. These values were used as error 
bars in Figure~\ref{fig:mod}.
}\label{tab_mod}
\begin{tabular}{@{}rrrrrrrrr@{}}
\hline
No & Freq & $\phi_b-\phi_y$ & $A_b/A_y$ & $\phi_v-\phi_y$ & $A_v/A_y$ & $\phi_B-\phi_V$ & $A_B/A_V$ & Average  \\ 
 & (d$^{-1}$) & (deg) &  & (deg) &  & (deg) &   & \\
\hline
$f_1$ & 13.360 & $-$4.13(-0.26) & 1.34 & $-$2.87(+1.0) & 1.523 & $-$4.61(-0.74) & 1.514 & -3.87   \\
$f_2$ & 10.476 & $-$4.22(-3.17) & 1.132 & +4.76(+5.81) & 1.044 & $-$1.59(-0.54) & 1.316 & -1.05   \\  
$f_3$ & 6.47 & $-$5.85(-0.24) & 1.038 & $-$0.80(+4.81) & 1.629 & $-$10.19(-4.58) & 1.827 & -5.61   \\
$f_4$ & 12.228 & $-$26.11(-3.84) & 1.077 & $-$18.44(+3.84) & 1.875 &  &  & -22.28    \\
$f_5$ & 11.614 & $-$15.26(-1.01) & 1.310 & $-$13.24(+1.01) & 1.840 & &  & -14.25  \\
$f_6$ & 6.916 & $-$6.12(-3.28) & 1.154 & +0.44(+3.28) & 1.114 & & & -2.84    \\
$f_8$ & 9.199 & $-$20.70(-7.47) & 0.904 & $-$5.26(+7.47) & 1.092 & $-$13.72(-0.49) & 2.013 & -13.23    \\
$f_{11}$ & 5.498 & $+$3.48(+0.46) & 1.022 & +15.92(+12.90)  & 1.709  & $-$10.33(-13.35) & 1.581 & +3.02   \\
$f_{13}$ & 9.604 & $-$22.59(+1.18) & 1.294 & $-$24.94(-1.18) & 1.864 &  & & -23.77   \\
$f_{15}$ & 5.813 & $+$2.48(-3.36)  & 0.577  & +11.81(+5.97)  & 0.890 & +3.23(-2.61) & 0.973 & +5.84  \\
$f_{18}$ & 10.204 & $+$4.93(+1.91) & 1.212 & +2.75(-0.26) & 1.261 & +1.39(-1.63) & 1.457 & +3.02    \\
\hline
\end{tabular}
\end{centering}
\end{table*}

The search for rotational splitting resulted in two doublets for $l=1$, two
incomplete multiplets for $l=2$, and two doublets for $l=3$. The $m$ values in 
Table~\ref{tab_rot} are tentatively given; there is no strong base for the 
identification, since we do not have complete triplets.
Averaging the observed spacing gives $\delta\nu_1=0.265$, $\delta\nu_2=0.439$, and 
$\delta\nu_3=0.486$. The three values are in good agreement with the 
theoretically expected values confirming that 38 Eri has a high inclination angle.
The observed ratios are $\delta\nu_1/\delta\nu_2=0.604$ and 
$\delta\nu_3/\delta\nu_2=1.107$. Although the limited number of frequencies 
(due to the short time base) did not allow us to get triplets and to identify the 
exact m values, but we concluded from the rotational splitting method 
that $l=1$, $l=2$, and $l=3$ modes are excited in 38 Eri according to this method. 
Since the frequencies of the highest amplitudes are included in the non-radial doublets/multiplets, 
except $f_1=13.360$, we may exclude that the $f_2=10.476$, $f_3=6.476$, and $f_4=12.228$~d$^{-1}$ 
frequencies are radial modes. The frequency at $f_1=13.360$~d$^{-1}$ could be 
a radial mode.  Checking the amplitude ratios in the doublets and the 
triplet, we found small amplitude for the possible $m=0$ and larger amplitude 
for the $m=\pm1$ frequencies. Similar amplitudes appeared in the triplet 
with $l=2$, as it was predicted by \citet{Gizon03} for high inclination.

However, a definite weakness of our result is that two
frequencies ($f_2$, and $f_3$) belong to 
doublets/multiplets with $l=1$ and $l=2$, or $l=2$ and $l=3$. Probably we 
missed some members of overlaping multiplets due to the small number of 
frequencies. In addition, the reviewer called our attention to the fact that two 
single spacings, $f_{14}$-$f_2$=0.341 and $f_7$-$f_{11}$=0.572~d$^{-1}$,
have 0.6 ratio. In this case the $P_{rot}$=1.466~d$^{-1}$ and the inclination is around 50 degree.  
However, no single spacing (0.625~d$^{-1}$) was found that could satisfy the other expected ratio. 
Although it is a charming possibility to find the inclination of the rotation from the spacings, 
but in my view the general behaviour of the rotation is not reflected by a single spacing. 
Nevertherless, observables resulted in the rotational splitting 
method for 38 Eri and the cases presented by \citet{Chen16,Chen17b,Chen17a} give a rather 
definite spacing range, presented in Table~\ref{tab_rs}. As \citet{Dziembowski92} 
stated, the g-mode asymptotic seems to work for the post main-sequence stars.

\begin{table}
\begin{centering}
\caption[]{The rotational split values for the post-main sequence stars, 
based partly on \citet{Chen16,Chen17a,Chen17b} and the present investigation.
}\label{tab_rs}
\begin{tabular}{@{}llll@{}}
\hline
Star & $\delta\nu_1$ & $\delta\nu_2$ & $\delta\nu_3$  \\
\hline
38 Eri & 0.265 & 0.439 & 0.486 \\
HD 50844 & 0.210 & 0.346 & 0.386 \\
EE Cam & 0.281 & 0.467 &  \\
CoRoT 102749568 & 0.384 & 0.643 & 0.706 \\ 
\hline
\end{tabular}
\end{centering}
\end{table}

\subsubsection{Sequence search}

We also applied the other approach, the sequence search method (SSA, \citealt{Paparo16a, Paparo16b}) 
for the 18 frequencies of 38 Eri. A visual inspection, concentrating on larger 
spacings, resulted in four sequences, each of them shifted to each other.
The sequences are given in Table~\ref{tab_ssa}. Sixteen frequencies are 
included in the four sequences out of the 18 {\it MOST} frequencies. There are two 
observables
that could be derived from the frequencies. The average spacing of each
sequence is given in the last column of Table~\ref{tab_ssa}. Averaging
these values, we get $\Delta\alpha=1.724\pm0.092$~d$^{-1}$. This is one of the 
observables of the SSA method.

Following the conclusion on the test case of FG Vir published in 
\citet{Paparo16a}, the $\Delta\alpha$ could be a combination of the large 
separation and the rotational frequency. In this case the sequences do not 
contain eigenmodes with the same $l$ value, but a mixture of eigenmodes 
and split modes due to the rotation, although we would like to have a
method for deriving the large separation for large amount of $\delta$ Scuti
stars presented by the space missions ({\it MOST}, {\it CoRoT}, {\it Kepler} and the forthcoming
{\it TESS} and {\it PLATO}). Of course, we can theoretically derive it if we do modelling
for each star, however, it is very time consuming. In our working hypothesis,
the comparison of the observables may yield to higher level regularities.
The numerical connection between the two methods is the following: 
$\Delta\alpha=6.53\cdot\delta\nu_1 = 3.95\cdot\delta\nu_2 = 3.58\cdot\delta\nu_3$. 
These relations do not show any higher level regularity, at least that we 
could easily explain from a single case. The co-efficients reflect the 
asymptotic rotational split ratio for different $l$ values. It would be worthwhile
to check for larger sample, whether the co-efficients are similar for each
cases or not. Similar values would suggest that the large spacing obtained by SSA and
the rotational split values are not independent of each other. At this moment 
we only conclude that the members of the rotatinal splitting belong to
different sequences. This means that the large spacing of 38 Eri is the
combination of the rotational frequency and the large separation.

Nonetheless, we used this
spacing and the scaling relation in \citet{Garcia Hernandez17}
to calculate the mean densities and surface gravities for 38 Eri. We
assumed that the spacing is just the large separation, since the
rotational splitting is relatively small. We aim to make an estimation of
these quantities, testing the validity of the scaling relation. The mean
density found in this way is $\bar\rho = [0.0394, 0.0554]$~gcm$^{-3}$. 
These are values obtained with the extreme cases of
the spacings (see Table~\ref{tab_ssa}). Using a large range in masses, 
$M = [1, 3]$~M$_{\sun}$, the estimated surface gravity is 
$\log g = [3.40, 3.66]$~cgs.
This is in agreement with the quantities given by \citet{Balona00} and
\citet{McDonald17}. Moreover, taken a mean value of the four spacings
and assuming a mass of $M = 2$~M$_{\sun}$, then $\log g = 3.53$, perfectly
matches Balona's and McDonald et al.'s values.

In the 
sequence search method not only is the (tentatively called) large separation  
an important parameter, but also the shift of the sequences to each other are. 
In the asymptotic regime of the non-rotating pulsating stars the sequence
of eigenmodes with different $l$ are systematically shifted to the radial modes
($l=0$). \citet{Paparo16b} proved for 90 $\delta$ Scuti stars that in 
non-asymptotic cases the shifts do not show a simple regularity, but we do 
not have any theoretical prediction for the shifts. However, the shift of the members of a
sequence to the members of the other sequence could be derived, as a new 
observable. For example, according to Table~\ref{tab_ssa} the shift of the II to I sequences are calculated as
the average of 5.813-5.498, 7.731-6.916, 11.614-10.476 and 13.360-12.228 
frequency differences. We compared the shifts, calculated forward (II-I) and 
backward (I-II) to the observed rotational split of 38 Eri. The ratios for both forward
and backward are given in Table~\ref{tab_shifts}.

\begin{table}
\begin{centering}
\caption[]{The ratios of the shift of the sequences to the observed
rotational split of 38 Eri. We calculated the ratios for both directions,
forward and backward. 
}\label{tab_shifts}
\begin{tabular}{@{}llllllll@{}}
\hline
Seq. & $\delta\nu_1$ & $\delta\nu_2$ & $\delta\nu_3$ & Seq. & $\delta\nu_1$ & $\delta\nu_2$ & $\delta\nu_3$   \\
\hline
II-I & 3.21 & 1.94 & 1.75 & I-II & 3.01 & 1.82 & 1.64 \\
IV-I & 2.65 & 1.60 & 1.44 & I-IV & 4.45 & 2.68 & 2.41 \\
III-I & 5.00 & 3.02 & 2.73 & I-III & 1.40 & 0.84 & 0.76 \\
IV-II & 3.69 & 2.23 & 2.01 & II-IV & 3.60 & 2.17 & 1.96 \\
III-IV & 3.04 & 1.83 & 1.66 & IV-III & 3.72 & 2.25 & 2.03 \\ 
III-II & 1.86 & 1.12 & 1.01 & II-III & 5.28 & 3.19 & 2.88 \\ 
\hline
\end{tabular}
\end{centering}
\end{table}

As Table~\ref{tab_shifts} shows, except the shift of IV-I and I-IV, all shifts are 
connected
by an integer times (2, 3, 5) to the rotational split of $l=1, 2$ and 3, but the
explanation needs more cases.
The two methods definitely show two different aspects of the 
underlying physical processes. 
For the completeness we checked the ratio of the rotational splits 
for i=50 degree to the shifts of the sequences. We have less cases with 
integer ratios, but we have some. The shift of IV-I is 2.06 times $\delta\nu_1$,
 I-IV is 2.06 times $\delta\nu_2$, III-I is 1.09 times $\delta\nu_1$, and 
II-III is 4.10 times $\delta\nu_1$. The integer values are also less precise.

An additional confirmation of the connection is supplied by 
{\it CoRoT} 102749568. The ratios of the rotational splitting values 
\citep{Chen17b}, presented in Table~\ref{tab_rs}. and
the independent shifts (table 6 of \citealt{Paparo13}) of the $l=0, 1$ and 2 sequences were
calculated.

The comparison of the rotational splitting and the shift of  the members of
the same sequences 
also resulted in an almost integer ratio. The differences between the same radial order 
of the $l=0$ and $l=1$ sequences were 2, 3 and 4 times the value of $\delta\nu_1$. 
More than a single ratio was obtained, since in this case we compared the 
independent shifts, not the averaged value, to the rotatinal splitting values. The consecutive
radial orders in the $l=0$ and $l=1$ sequences differ by two times $\delta\nu_3$. 
The consecutive radial orders of the $l=2$ and $l=0$ sequences differ by five times 
$\delta\nu_1$ or three times $\delta\nu_2$.
   It seems that pulsation and rotation have a strong effect on each other that
it is not easy to disentangle, but maybe these methods take us closer to the solution.

\section{Mode identification}\label{sec:iden}

\subsection{Period ratio and pulsation constant}

Since the systematic theoretical calculation of the period ratios
\citep{Stellingwerf79} and the pulsation constants \citep{Fitch81} for $\delta$
Scuti stars, both observables were widely used in the 80s for mode 
identification (see \citealt*{Breger79,Poretti87,Poretti88,Poretti89}).
In radially pulsating stars the period ratios of the excited modes are good
tools for mode identification. The ratio of the radial overtones yield
certain values that we can use for mode identification. The theoretical radial 
period ratios for the non-radially pulsating $\delta$ Scuti stars 
calculated by \citet{Stellingwerf79} are 0.756-0.789 for $P_1/P_0$, 
0.611-0.632 for $P_2/P_0$ and 0.500-0.525 for $P_3/P_0$, 
where $P_0$, $P_1$, $P_2$ and $P_3$ are 
periods of the radial fundamental, first, second and third radial overtone modes.  
\begin{figure*}
\includegraphics[angle=0,scale=0.61]{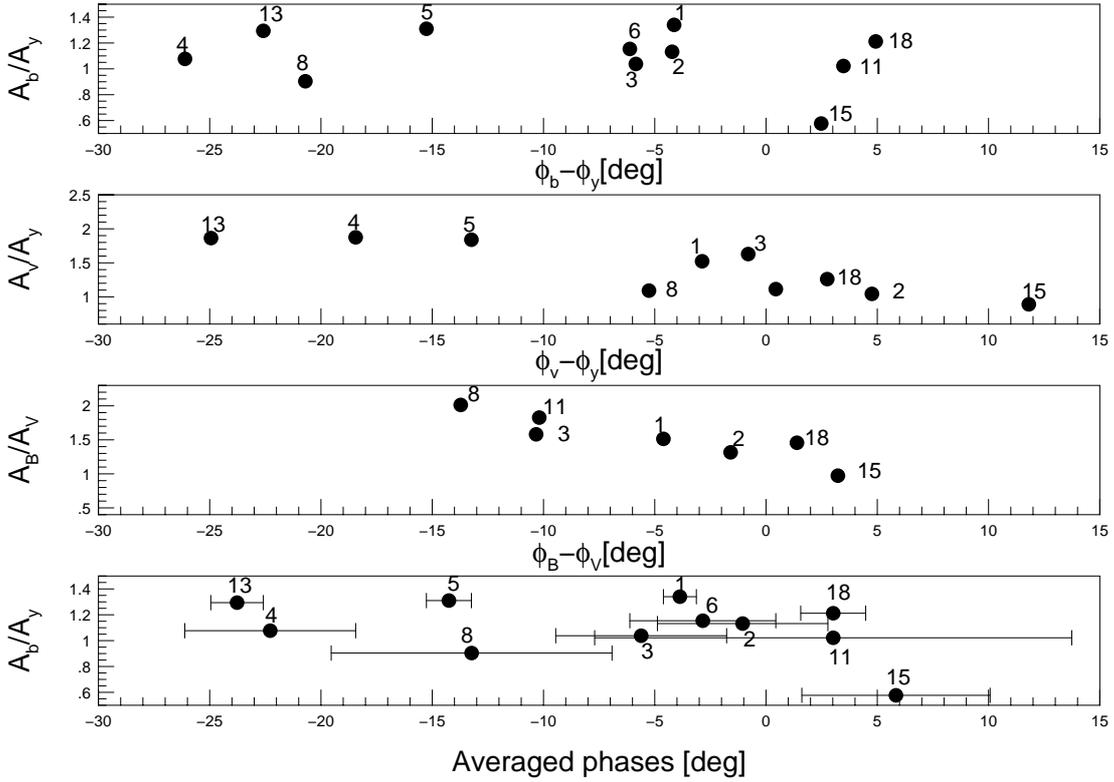}
\caption{
Discriminative panels for mode identification on 
Str\"omgren {\it b, v, y}, Johnson {\it B} and {\it V} colours. 
In the fourth panel the average
of the phase differences of the previous panels is shown versus the amplitude ratios of
the first panel. Each
amplitude ratios and phase differences are given in Table~\ref{tab_mod}. The deviation
of the phase differences from the averaged value are given in bracket.
Frequences are separated in three different groups especially in 
the top panel. The location of the same
frequencies can be followed from panel to panel at the same place. An
estimation of error in the phase differences were made using the three two-colour 
diagrams. The frequencies are well-separated in phase differences in the fourth panel.
The indentification of frequencies are based on the general trend of the
theoretical investigation, namely, that the phase differences change from right to left
for the $l=0, 1$ and 2 modes. 
The identification obtained in our hybrid methods is given in Table~\ref{tab_comp_mod}.
}
\label{fig:mod}
\end{figure*}

The subsequent calculations on the dependence of the period ratios on
the metallicity and the rotation \citep*{Suarez06a,Suarez06b,Suarez07} 
revealed that both the higher metallicity and the higher
rotational velocity increase the period ratios with around $10^{-3}$. The
typical $P_1/P_0$ values found for main sequence Pop I stars are in the
range of [0.772, 0.776]. Unfortunately, there is a mix-up between the two
effects, so the usefulness of the period ratios for mode identification is
not as straightforward as it was thought in the early 80s.
In the lack of new period ratios for higher radial overtones, we used the 
rather wide range of \citep{Stellingwerf79} for checking the modes of 38 Eri.

The $f_3/f_2=0.618$ ratio suggests that  
$f_3=6.477$~d$^{-1}$ is the radial fundamental and $f_2=10.476$~d$^{-1}$ is 
the radial second overtone. 
Accepting $f_3=6.477$~d$^{-1}$ as a radial fundamental mode, 
the $f_3/f_1=0.485$ and $f_3/f_4=0.530$
ratios rule out that either $f_1$ or $f_4$ would be the third radial overtone.
However, the $f_2/f_1=0.784$ ratio suggests that $f_2=10.476$~d$^{-1}$ is a 
radial fundamental and $f_1=13.360$~d$^{-1}$ is the radial first overtone
according to Stellingwerf's wider range, however, the more severe 
[0.772, 0.776] range of \citet{Suarez06a} rules out that $f_1$ and $f_2$ would 
be radial modes.

We calculated the $Q$ value for the four frequencies, 
$f_1$, $f_2$, $f_3$ and $f_4$, using the empirical relation 
\begin{equation}
   \log Q = \log P+\frac{1}{2}\log g + \frac{1}{10}M_{\mathrm{bol}} + \log T_{\mathrm{eff}} -6.454
\end{equation}
given by \citet{Bowman16} and  both sets of the physical parameters from Table~\ref{tab_phys}.
The typical values of pulsation constants for fundamental, first and second 
overtone radial p-modes in $\delta$ Scuti stars lie
in the range $0.022\le Q\le 0.033$ \citep{Breger75}. 
The empirical relation resulted in $Q=0.032$  or 0.029 for $f_3=6.477$~d$^{-1}$ 
allowing the possibility for being a radial mode. All the other 
frequencies of highest amplitude are out of the range.
A further check resulted in $Q=0.020$ or $0.019$ for $f_{15}$ and $Q=0.033$ 
for $f_{18}$ with McDonald et al's parameters. These values are at the boarders of the
range for radial modes. We conclude that neither period ratios, nor pulsation
constants gave well-established mode identification for the frequencies of 38 Eri.

\subsection{Multi-colour photometry}

\citet{Dziembowski77} derived that the amplitude and phase values in different 
passbands can distinguish the frequencies with different horizontal quantum number ($l$). 
\citet{Watson88} published the discriminative boxes in 
Johnson {\it B} and {\it V} colours, exactly for $A_{B-V}/A_V$ versus $\phi_{B-V}-\phi_V$. 
Although the theoretical location of the frequencies with $l=1$ and $l=2$ partly 
overlap, this method was widely used in the past for mode identification. 
\citet*{Garrido90}, based on linear approximation, published the modal discrimination 
in pulsating stars by using Str\"omgren photometry. The theoretically expected location 
of the frequencies for $l=0,1$ and 2 were mostly overlaping for different
filter combinations and the calculations were restricted to the radial
pulsation constant ($Q=0.033$). Only
the $A_{b-y}/A_y$ amplitude ratio versus
 $\phi_{b-y}-\phi_y$ proved to be discriminative enough to compare it to the
observed values. In this approach the $\phi_{b-y}-\phi_y$ phase difference is
positive, while the other modes show negative phase differences. The
later investigations using non-adiabatic pulsational treatment of the atmoshpere 
\citep*{Moya04,Daszynska-Daszkiewicz03} revealed that the results are
highly sensitive to the convection, i.e. to the mixing length parameter ($\alpha$).
Even negative phase differences were resulted for the higher overtones
($Q=0.017$) of the $l=0$ modes.
Nevertheless, the phase difference values for 
frequencies with higher $l$ are always displayed from right to left, 
disregarding the exact values.
Different approaches are followed in practice. According to \citet{Balona99}, 
the best approach to make identification is 
to make use of all available information. A $\chi^2$ goodness-fit criteria 
between observed amplitudes and amplitudes predicted for a given $l$ over
as many wavebands as desired were used for mode identification. The method
avoids the use of numerous two-colour diagrams. In one of the latest example
\citep{Breger17}, instead of the color index value ($\phi_{v-y}-\phi_y$) the
colour phase differences ($A_v/A_y$ versus $\phi_v-\phi_y$) were used for
mode identification. The amplitudes and phases are more precisely determined
from the colour light curves of highest amplitudes than from the colour index 
curves. The aforementioned examples show the difficulty of the mode 
identification.

We tried to use what we have for the mode identification in 38 Eri. We have 
amplitudes and phases in Johnson {\it B} and {\it V} and Str\"omgren {\it b, v, y} colours obtained 
in 1998. We have precise frequencies obtained from the {\it MOST}
data from 2011. Although the frequencies used to be determined from the data set 
that is used for mode identification, even the multi-site ground-based campaign 
was not adequate enough to get the same  well-determined frequency solution 
for the colours. At the same time the {\it MOST} observations were not obtained in different filters. 
We used a hybrid solution. We determined the color amplitudes and phases (relative to 
the HJD$=$2\,541\,113.0 epoch) using the {\it MOST} frequencies (see in Table~\ref{tab_fr_MOST}). 
In some cases too high amplitudes or too large phase values were obtained and thus omitted.
Nevertheless, we found reasonable amplitudes and phases for 11 {\it MOST} frequencies. 
The amplitude ratios and phase differences are 
given in Table~\ref{tab_mod} for Str\"omgren colours {\it b} and {\it v}, and {\it y} 
and Johnson {\it B} and 
{\it V}. Although we calculated the amplitude ratios and phase differences for the 
{\it b$-$y}, {\it v$-$y} and {\it B$-$V}  colour indices, too, but due to the much larger error 
connected to the much lower amplitude and the additional mathematical 
calculation, the solutions were not as conclusive as 
for the colours. We followed \citet{Breger17} in using the colour values 
instead of the color index values. 
Unfortunately, we do not have such a previleged situation as 
\citet{Breger17} had for 4 CVn. They calculated the amplitude and phase values 
from season to season
and presented the scatter as an error. We had to find an hybrid solution, too,
for the error calculation in the case of 38 Eri. At first we tried the Monte 
Carlo method for error estimation, however, it revealed unrealistic errors,
changing from 5-20 degrees for the phase 
differences and 0.07-0.3 for the amplitude ratios. This was due to the application of 
{\it MOST} frequencies for the colour data. However, we can use the advantage of having 
observations in different passbands. Different combination of the two-colour
diagrams are shown in the different panels of Figure~\ref{fig:mod} displaying similar
arrangement of the frequencies. 
The first panel showing $A_b/A_y$ versus $\phi_b-\phi_y$ is the
most conclusive. There are three well-separated groups, although the 
theoretical investigations do not predict such a nice separation
for $l=0, 1$ and 2 modes \citep{Garrido90,Moya04}.
The number beside the dots
gives the ID of frequencies from Table~\ref{tab_fr_MOST}. Although in the other panels the
groups are not as clearly separated, the numbers help to follow how the
location of a certain frequency changes from panel to panel.

A more reliable error estimation would
be to average the phases of the three different colour relations given in
Table~\ref{tab_mod} and use the scatter as the error. Although the theory
suggests different phase differences for the different colour relations, but
we can get an upper limit for the error. The averages and the scatters as 
errors are presented also in Table~\ref{tab_mod} and are displayed in the
fourth panel of Figure~\ref{fig:mod}. As the amplitude ratios are remarkably
different in the colours, we used the $A_b/A_y$ amplitude ratios only
for the presentation as a hybrid solution, without any additional meaning.

Supposing that the multi-colour photomery with the {\it MOST} frequencies gives
reasonable approximation for the amplitude ratios and phase differences, we 
used them for mode identification given
in the last four columns of Table~\ref{tab_comp_mod}. As the location
of the modes with different $l$ are contradictory according to the theoretical
investigations using different assumptions, we used only the general
trend that is common in each investigation. The phase differences change
from right to left for the $l=0, 1, 2$ and 3 modes. We accepted the identification for $f_1$,
$f_2$, $f_3$ and $f_6$ as $l=1$ modes, for $f_5$, and $f_8$, as $l=2$ and 
 $f_4$ and $f_{13}$ as $l=3$ and $f_{15}$ and $f_{18}$ as $l=0$ radial modes. 
We do not accept $f_{11}$ as an $l=0$ mode due to the large error bar.
Although $f_{15}$ and $f_{18}$ are in the positive region the frequency ratio 
do not fit the regions given by \citet{Stellingwerf79} for the radial modes.

For comparison the third column gives the mode identification 
according to the rotational splitting. 
There is no agreement between the identification by the rotational 
splitting and the finally accepted identification of the multicolour photometry. 
The fourth column gives the numbering of sequences
according to the sequence search method. Each sequence contains frequencies 
with all $l$ values. Taking the fast rotation of 38 Eri into 
account, the deviation of the real sequences from the 
sequences of the pure eigenmodes \citep{Paparo16a} is also not surprising.
 
\begin{table}
\begin{centering}
\caption[]{Comparison of the identification with different methods: 
rotational splitting, sequence search (SSA) and multi-colour photometry. 
For multi-colour photomery identifications on different colour 
combinations and on the averaged phase differences are given.
}\label{tab_comp_mod}
\begin{tabular}{@{}rrrccccc@{}}
\hline
    &            & Rot. sp. &  SSA  &  \multicolumn{4}{c}{Multi-colour phot.}  \\
    &            &          &       &  {\it b, y} & {\it v, y} & {\it B, V} & Average  \\
ID  & Freq.      & $l$      & No    &  $l$        & $l$        & $l$       & $l$   \\ 
    & (d$^{-1}$) &          &       &             &            &           &        \\
\hline
$f_1$ & 13.360 & 0  & II &  1 & 1 & 1 & 1  \\
$f_2$ & 10.476 & 1/2 &   I & 1 & 0 & 1 & 1  \\  
$f_3$ & 6.477 & 2/3 &   III & 1 & 1 & 1 & 1  \\
$f_4$ & 12.228 &  &  I  & 2 & 2 &  & 3   \\
$f_5$ & 11.614 &  &  II &  2 & 2 & - & 2  \\
$f_6$ & 6.916 & 2 &   I &  1 & 0 &  - & 1  \\
$f_7$ & 6.070 & 1 &  IV &  - & - & -  & - \\
$f_8$ & 9.199 &  &   IV &  2 & 1 & 2  & 2  \\
$f_9$ & 7.033 & 3 &  - &  - & - & - & -    \\
$f_{11}$ & 5.498 & 3 &   I &  0 & 0 & 1 & -  \\
$f_{12}$ & 7.731 & 2 &  II & - & - & - & -  \\
$f_{13}$ & 9.604 & 2  &  II & 2 & 2 & - & 3   \\
$f_{14}$ & 10.817 &  &  IV & - & - & - & -   \\
$f_{15}$ & 5.813 & 1 &  II & 0 & 0 & 0 & 0  \\
$f_{16}$ & 13.966 &  &   I & - & - & - & -   \\
$f_{17}$ & 11.827 &  &  III & - & - & - & -  \\
$f_{18}$ & 10.204 & 1 &  III &  0 & 0 & 0 & 0  \\
\hline
\end{tabular}
\end{centering}
\end{table}

Mode identification using {\sc Famias} \citep{Zima08a, Zima08b} were carried out for comparing 
the theoretical values to our observational values. Using  
$T_{\mathrm{eff}}=7100\pm120$~K, $\log g=3.6\pm0.06$ and the standard assumptions 
for $\delta$ Scuti stars  (Kurucz model, no overshooting, solar metallicity
and the model of \citealt{Montalban07}), the best agreements are $l=1$ for $f_1$ and $l=2$ for 
$f_5$ and $f_8$  which agree with results of the identification of our hybrid method. 
Definitely modelling by a code for fast rotating
$\delta$ Scuti stars (e.g. \citealt{Lignieres09, Reese17}) would be more appropriate for 
38 Eri.

\section{Summary}\label{sec:concl}

The present investigation of the $\delta$ Scuti star 38 Eri resulted in a significant 
improvement in the description of the pulsational behaviour. We resolved the 
different causes of the alias discrepancies presented in previous 
investigations. We determined the frequency content at two  
epochs separated by 13 years, suggesting frequencies changing amplitude for the 
second epoch.  Neither frequencies at 5.2 and 6.2~d$^{-1}$, obtained in 
the colour data, appeared in the {\it MOST} data, but frequencies at 6.477~d$^{-1}$ ($f_3$) and 
5.498~d$^{-1}$ ($f_{11}$) were found. Although both $f_2=10.476$ and $f_{18}=10.204$~d$^{-1}$
appeared in the {\it MOST} data, however, a frequency at 10.214~d$^{-1}$ had the second
highest in Johnson {\it B} and {\it V} colours.
It is not known whether these frequencies 
are different members of the same rotational split triplets showing 
different amplitude at different epoch or not.

We found only amplitude variability and only for a single frequency, 
$f_3=6.477$~d$^{-1}$. We excluded beating of close pairs and resonant 
mode coupling as possible cause of the amplitude variation. 
Following \citet{Bowman16} we concluded that 
variable driving/damping could be responsible for the amplitude variation.

We applied 
two methods for finding spacings: the rotational splitting method for the 
small spacings and the sequence search method for the larger spacings.
We proved by modelling that the asymptotic relation for getting the rotational
splittings can be used for 38 Eri. Using the scaling relation we calculated the
mean density $\bar\rho =[0.0394, 0.0554]$~gcm$^{-3}$ and surface gravity 
$\log g = [3.40, 3.66]$~cgs from the spacing obtained in the sequence search method.
At the first time we compared the observables which can be obtained from
these methods. The
two methods seem to be numerically connected. The ratios of some shifts 
between the sequences and the rotational splitting found for $l=1, 2$ and 3 are
resulted in integer values (2, 3, 5). We presented similar ratios for the $\delta$ Scuti
star, CoRoT 102749568 as a support for our present result.

We emphashized the difficulty of the mode identification for $\delta$ Scuti stars, 
but we followed a positive attitude for finding some solution
based on our large efforts of the ground-based multi-site multi-colour 
photometry and the {\it MOST} observations.
We used a hybrid method for mode identification in two senses. Partly we combined 
the precise frequencies obtained on {\it MOST} data in 2011 and the extended ground-based 
multi-colour photometry from 1998. The
acceptable results for the phase difference and period ratios show the capability 
of the hybrid method. Obtaining additional multi-colour photometry for the frequencies 
of high precision for the tremendous number of {\it CoRoT} and {\it Kepler} $\delta$ Scuti stars, 
we could  provide a great step towards the identification of modes in the non-asymptotic regime 
of the pulsation.
Secondly, we used an unusual way of getting the final mode identification based on the 
multi-colour photometry. We present the discriminative panels of the 
different colour combinations separately, however, we stepped further. 
We calculated the averaged phase differences of the different color combinations and used 
the scatter of the independent values to the average as an error. We
used for mode identification the general trends of the theoretical calculations 
instead of certain boxes derived. The phase differences change from right to
left for the $l=0, 1, 2$, and 3 modes. This method resulted in two modes with $l=0$ 
($f_{15}$ and $f_{18}$). A third frequency ($f_{11}$) resulted also in a positive
phase difference, but with a large error bar. Four modes ($f_1$, $f_2$, $f_3$ and 
$f_6$) located in the low negative region are identified as $l=1$ modes.
Two well-separated higher phase difference values appear to suggest an 
identification with $l=2$ for $f_5$ and $f_8$, and with $l=3$ for $f_{13}$ and 
$f_4$. {\sc Famias} resulted in the same identification for $f_1$, $f_5$ and $f_8$.

There is no agreement between the identification based on the rotational 
splitting and the multi-colour photometry. The sequences do not contain  
modes with the same $l$ values. At this moment we may not conclude that we
can have mode identification for $\delta$ Scuti stars based only on the
frequencies. However, 38 Eri is only one example. We have precise frequencies
for plenty of $\delta$ Scuti stars by the {\it CoRoT} and {\it Kepler} space missions and
we will have even more by the forthcoming {\it TESS} and {\it PLATO} missions. We believe
that new approaches are needed for getting the information that are at our 
hands in the space data. Maybe, the frequencies of even such a 
complex $\delta$ Scuti star as 38 Eri could be successfully identified in 
space data with longer time base,
especially if the theoretical mode identification is also applied \citep{Reese17}.
  
\section{Acknowledgements}

MP was supported by Soros Foundation. She also thanks the staff of SAAO. 
MP thanks Gerald Handler for his remarks and comments on behalf of the 
late Robert Shobbrook. ZK observed in the frame of the exchange agreement between the Hungarian
 Academy of Sciences and CONACYT.This work was supported by the ESA PECS Grant No
4000103541/11/NL/KML.  
\'AS was supported by the J\'anos Bolyai Research Scholarship of the
Hungarian Academy of Sciences, and he also acknowledges the financial
support of the Hungarian NKFIH Grant K-113117. 
Spectroscopic observations made with the Mercator Telescope, operated on
the island of La Palma by the Flemmish Community, at the Spanish
Observatorio del Roque de los Muchachos of the Instituto de
Astrof\'{i}sica de Canarias. \'AS, JMB and
ZsB acknowledge the financial support of the Hungarian NKFIH Grants
K-115709 and K-119517. ZsB acknowledges the support provided from the
National Research, Development and Innovation Fund of Hungary,
financed under the PD\_17 funding scheme, project No. PD-123910.
AK acknowledges the Science and Education Ministry 
of Kazakhstan (grant No. 0075/GF4). 
AFJM is grateful for financial aid from NSERC (Canada) and FQRNT (Quebec).
AGH acknowledges funding support from Spanish public
funds for research under project ESP2015-65712-C5-5-R (MINECO/FEDER), and
from project RYC-2012-09913 under the `Ram\'on y Cajal' programme of the
Spanish MINECO.
We acknowledge the International Space Science Institute (ISSI) for
supporting the SoFAR international
team.\footnote{\url{http://www.issi.unibe.ch/teams/sofar/}}
We thank to the anonymous referee for the very thorough work.





\appendix

\subsection*{Affiliations}

{\footnotesize \it{
$^{1}$Konkoly Observatory, MTA CSFK, Konkoly Thege M. u. 15-17., H-1121 Budapest, Hungary\\
$^{2}$E\"otv\"os Lor\'and University, Savaria Department of Physics, K\'arolyi G. t\'er 4., H-9700 Szombathely, Hungary\\ 
$^{3}$Australian National University, Siding Spring Observatory, Coonabarabran, NSW 2137, Australia\\
$^{4}$Department of Physics and Astronomy, University of British Columbia, 6224 Agricultural Road, Vancouver, BC V6T 1Z1, Canada \\
$^{5}$Stellar Astrophysics Centre, Department of Physics and Astronomy, Aarhus University. 120 Ny Munkegade, DK-8000 Aarhus C, Denmark \\
$^{6}$Korea Astronomy and Space Science Institute, 39-18 Hwaam-dong, Yuseong-gu, Daejeon 34055, Republic of Korea\\
$^{7}$Avcorp Industries, Inc. 10025 River Way, Delta, BC, V4G 1M7, Canada\\
$^{8}$Department of Physics, Alzahra University, P.O. Box 1993893973, Tehran, Iran \\
$^{9}$8100 Barstow St. NE, Apt 6104, Albuquerque, NM, 87122, USA\\
$^{10}$Fesenkov Astrophysical Institute, Observatory 23, 050020, Almaty, Kazakhstan\\
$^{11}$Department of Theoretical Physics and Cosmology,
University of Granada (UGR), E-18071 Granada, Spain\\
$^{12}$Instituto de Astronom\'{\i}a, UNAM, 
 Apartado Postal 70-264, Ciudad M\'exico, CDMX, C.~P. 04510, Mexico \\ 
$^{13}$Institut f\"ur Astronomie, Universit\"at Wien, T\"urkenschanzstrasse 17, A-1180 Wien, Austria \\
$^{14}$Institut f\"ur Kommunikationsnetze und Satellitenkommunikation, Technische
Universit\"at Graz, Infeldgasse 12, 8010 Graz, Austria\\
$^{15}$D\'epartement de physique and Centre de Recherche en Astrophysique du
Qu\'ebec (CRAQ), Universit\'e de Montr\'eal, C.P. 6128, 
Succ. Centre-Ville, Montr\'eal, QC H3C 3J7, Canada\\
$^{16}$Canadian Coast Guard College, Dept. of Arts, Sciences, and Languages, Sydney, Nova Scotia, B1R 2J6, Canada\\
$^{17}$Department of Astronomy and Astrophysics, University of Toronto, Toronto, ON M5S 3H4, Canada\\
$^{18}$International Centre for Radio Astronomy Research, University of Western Australia 
35 Stirling Hwy, Crawley, WA 6009, Australia\\
}
}

\bsp	
\label{lastpage}
\end{document}